\documentclass[fleqn,12pt]{wlpeerj}	% I changed it to 12 from 10 as given in website

\usepackage[normalem]{ulem}
\usepackage[detect-all]{siunitx}
\usepackage{dcolumn}
\usepackage{graphicx}
\usepackage{adjustbox}
\usepackage{multirow}
\usepackage{booktabs}
\usepackage{listings} % to include c code
\usepackage{relsize}
\usepackage{amsmath}
\usepackage[T1]{fontenc}
\usepackage{times}%
\usepackage{url}
\usepackage{amsmath}
\usepackage{hyperref}
\usepackage{etoolbox}% http://ctan.org/pkg/etoolbox
\AtBeginEnvironment{align}{\setcounter{subeqn}{0}}% Reset subequation number at start of align
\newcounter{subeqn} \renewcommand{\thesubeqn}{\theequation\alph{subeqn}}%
\newcommand{\subeqn}{%
 ~\refstepcounter{subeqn}% Step subequation number
  \tag{\thesubeqn}% Label equation
}

\usepackage{dcolumn}
\usepackage{textcomp}
\usepackage{mathtools}

\definecolor{myblue}{rgb}{0, 0, 0} % {\color{myblue}}
\definecolor{myblue1}{rgb}{0, 0, 0} % {\color{myblue}} % 0.06, 0.3, 0.57

\definecolor{myyaleblue}{rgb}{0, 0, 0} % review 1
\definecolor{myrosewood}{rgb}{0, 0, 0} % review 2
\definecolor{myviolet}{rgb}{0, 0, 0} % review 3
\definecolor{myred}{rgb}{0.6, 0.03, 0.07} % {\color{myblue}}
\usepackage[title]{appendix}
\usepackage{url}
\usepackage{enumitem}	% for paragraph change in enumerate
\setlist{parsep=0pt,listparindent=\parindent}

\usepackage{algpseudocode}
\usepackage{algorithmicx}
\RequirePackage{algorithm}
\usepackage{romannum}	% for Romun number

\usepackage{multicol} % multicol in Algorithm
\usepackage{amsmath}

\DeclareMathOperator*{\argmin}{argmin}
\usepackage{subcaption}
\usepackage{hyperref}
\newcommand\hati{\hat\imath}% MCS: \hat{i} is usually considered poor style.
\usepackage{makecell} % for think line use \Xhline{2\arrayrulewidth} instead of \hline
\usepackage{footmisc}

\title{SABMIS: Sparse Approximation based Blind Multi-Image Steganography Scheme}

\author[1]{Rohit Agrawal}
\author[1]{Kapil Ahuja}
\author[2]{Marc C. Steinbach}
\author[2]{Thomas Wick}

\affil[1]{Computer Science and Engineering,
Indian Institute of Technology Indore, India}

\affil[2]{Leibniz Universit\"at Hannover,
  Institut f\"ur Angewandte Mathematik,
  Hannover, Germany}

\corrauthor{Kapil Ahuja}{kahuja@iiti.ac.in}

\begin{abstract}
\fontsize{10pt}{12pt}\selectfont
{
We hide grayscale secret images into a grayscale cover image, which is considered to be a challenging steganography problem. Our goal is to develop a steganography scheme with enhanced embedding capacity while preserving the visual quality of the stego-image as well as the extracted secret image, and ensuring that the stego-image is resistant to steganographic attacks.

\hspace{0.5cm}
The novel embedding rule of our scheme helps to {\color{myyaleblue}hide secret image sparse coefficients into the oversampled cover image sparse coefficients in a staggered manner.} The stego-image is constructed by using the Alternating Direction Method of Multipliers (ADMM) to solve the Least Absolute Shrinkage and Selection Operator (LASSO) formulation of the underlying minimization problem. Finally, the secret images are extracted from the constructed stego-image using the reverse of our embedding rule. {\color{myyaleblue}Using these components together, to achieve the above mentioned competing goals, forms our most novel contribution}. We term our scheme SABMIS (Sparse Approximation Blind Multi-Image Steganography).

\hspace{0.5cm}
{\color{myyaleblue}We perform extensive experiments on several standard images. By choosing the size of the length and the width of the secret images to be half of the length and the width of cover image, respectively, we obtain embedding capacities of \SI{2} bpp (bits per pixel), \SI{4} bpp, \SI{6} bpp, and \SI{8} bpp while embedding one, two, three, and four secret images, respectively. Our focus is on hiding multiple secret images. For the case of hiding two and three secret images, our embedding capacities are higher than all the embedding capacities obtained in the literature until now (3 times and 6 times than the existing best, respectively). For the case of hiding four secret images, although our capacity is slightly lower than \citep{Ste_Ios_2006} (about ${\frac{2}{3}}^{rd}$), we do better on the other two goals (quality of stego-image \& extracted secret image as well as resistant to steganographic attacks).} 

\hspace{0.5cm}
For our experiments,  there is very little deterioration in the quality of the stego-images as compared to their corresponding cover images. {\color{myyaleblue}Like all other competing works, this is supported visually as well as over \SI{30} dB of Peak Signal-to-Noise Ratio (PSNR) values. The good quality of the stego-images is further validated by multiple numerical measures. None of the existing works perform this exhaustive validation. When using SABMIS, the quality of the extracted secret images is almost same as that of the corresponding original secret images. This aspect is also not demonstrated in all competing literature.}

\hspace{0.5cm}
 {\color{myyaleblue}SABMIS further improves the security of the inherently steganographic attack resistant transform based schemes. Thus, it one of the most secure schemes among the existing ones. Additionally, we demonstrate that SABMIS executes in few minutes, and show its application on the real-life problems of securely transmitting medical images over the internet.}
 }
\end{abstract}

\begin{document}
\maketitle
\flushbottom
\thispagestyle{empty}

\section{Introduction}
The primary concern during the transmission of digital data over communication media is that anybody can access this data. Hence, to protect the data from being accessed by illegitimate users, the sender must employ some security mechanisms. In general, there are two main approaches used to protect secret data; cryptography \citep{Stalling} and steganography \citep{PeerjImagSte1}, with our focus on the latter. %\citep{Gutub2021}

Steganography is derived from the Greek words \emph{steganos} for ``covered'' or ``secret'' and \emph{graphie} for ``writing''. In steganography, the secret data is hidden in some unsuspected cover media so that it is visually imperceptible. Here, both the secret data as well as the cover media may be text or multimedia. Recently, steganography schemes that use images (binary, grayscale or color) as secret data as well as cover media have gained a lot of research interest due to their heavy use in World Wide Web applications. This is the \emph{first} focus of our work \footnote{{\color{myyaleblue}Hiding binary data into images is a different track, which we are not focusing in this paper. For the sake of completeness, this is summarized in Appendix \ref{appendix:Some steganography schemes for hiding binary secret data}.}}.
{\color{myyaleblue}Some real-life applications of this include securing biometric data, digital signature, personal banking information, and medical data.}

Next, we present some relevant previous studies in this domain. Secret data can be hidden in images in two ways; spatially or by using a transform. In the spatial domain based image steganography scheme, secret data is hidden directly into the image by some modification in the values of the image pixels. {\color{myrosewood}These approaches have the drawback that they are inherently not resistant to steganographic attacks \citep{LSB_Not_Resistant_MDPI,  LSB_Not_Resistant_IEEETra}.} Some of the past works related to this are given in Table \ref{Table:Spatial domain based image steganography schemes}. The papers in this table are listed in the increasing order of the number of secret images hidden in the cover image.

In the transform domain based image steganography scheme, first, the image is transformed into frequency components, and then the secret data is hidden into these components. {\color{myrosewood}This process makes these approaches intrinsically resistant to steganographic attacks. Hence, such approaches form our \emph{second} focus.} Some of the past works related to this are given in Table \ref{Table:Transform domain based image steganography schemes}. The papers in this table are listed in the increasing order of the number of secret images hidden in the cover image as well.

%\mcs{MCS: maybe drop Sr.~No. (entire left column)
%  and shorten text in two rightmost columns as indicated.}

\begin{table}[!h]
%\def\?#1 {\mcs{\sout{#1}} }% Note: trailing space is essential!
%{\color{myyaleblue}
\centering
\footnotesize
\caption{Spatial domain-based image steganography schemes.}
\label{Table:Spatial domain based image steganography schemes}
\setlength{\tabcolsep}{2.5pt}
\begin{tabular}{|c|c|c|c|}
\hline
\textbf{Reference} & \textbf{Technique}                                                          & \textbf{\begin{tabular}[c]{@{}c@{}}Secret images\end{tabular}}    & \textbf{\begin{tabular}[c]{@{}c@{}}Cover image \end{tabular}}                                                          \\ \hline
 \citep{IEEE_Ste2019} & \begin{tabular}[c]{@{}c@{}}A modified version of \\ Least Significant Bits\\ (LSB) with deep \\ neural networks\end{tabular} & 2 color & \begin{tabular}[c]{@{}c@{}}color \end{tabular} \\ \hline
 \citep{Ste_2020} & \begin{tabular}[c]{@{}c@{}}LSB\end{tabular}  & 2 color  & color                                                           \\ \hline
 \citep{Prashanti} & \begin{tabular}[c]{@{}c@{}}LSB\end{tabular} & 3 binary & \begin{tabular}[c]{@{}c@{}}grayscale and color \end{tabular} \\ \hline
\citep{Ste_Ios_2006} & \begin{tabular}[c]{@{}c@{}}A modified version of \\ LSB\end{tabular} & 4 grayscale &  \begin{tabular}[c]{@{}c@{}}grayscale \end{tabular} \\ \hline
\citep{Ste_2014_4hide} & \begin{tabular}[c]{@{}c@{}}A modified version of \\ LSB\end{tabular} & 4 color &  \begin{tabular}[c]{@{}c@{}}color \end{tabular} \\ \hline 
\end{tabular}
%}
\end{table}

\begin{table}[!h]
%\def\?#1 {\mcs{\sout{#1}} }% Note: trailing space is essential!
%{\color{myyaleblue}
\centering
\footnotesize
\caption{Transform domain-based image steganography schemes.}
\label{Table:Transform domain based image steganography schemes}
\setlength{\tabcolsep}{2.5pt}
\begin{tabular}{|c|c|c|c|}
\hline
 \textbf{Reference}                                 & \textbf{Technique}                                                                                                                         & \textbf{\begin{tabular}[c]{@{}c@{}}Secret images  \end{tabular}}    & \textbf{\begin{tabular}[c]{@{}c@{}}Cover image \end{tabular}}                                                          \\ \hline
 \citep{Sanjutha_MK}                          & \begin{tabular}[c]{@{}c@{}}Discrete Wavelet \\ Transformation (DWT)\\ with Particle Swarm \\ Optimization (PSO)\end{tabular}                                              & 1 grayscale  & color     \\ \hline
 \citep{S_Arunkumar_MedicalImage_JIFS}      & \begin{tabular}[c]{@{}c@{}}Redundant Integer Wavelet \\ Transform (RIWT) and QR \\ Factorization\end{tabular}                      & 1 grayscale     & color     \\ \hline
\citep{MMTA_QRSteg} & \begin{tabular}[c]{@{}c@{}}Contourlet and Fresnelet \\ Transformations with \\ Genetic Algorithm (GA) \\ and PSO\end{tabular} & \begin{tabular}[c]{@{}c@{}}1 binary \\(specifically, QR code) \end{tabular}  & grayscale \\ \hline
\citep{S_Arunkumar_MedicalImage_DWT_2019} & \begin{tabular}[c]{@{}c@{}}RIWT, Singular Value \\ Decomposition (SVD), \\ and Discrete Cosine \\ Transformation (DCT)\end{tabular} & 1 grayscale   & grayscale \\ \hline
 \citep{S_Hemalatha}                          & \begin{tabular}[c]{@{}c@{}}DWT\end{tabular}                                                   & 2 grayscale & color      \\ \hline
 \citep{Ste_2020}                             & DWT and SVD                                                                                                                        & 2 color       & color     \\ \hline
% \citep{Gutub2021}                             & DWT                                                                                                                     & binary data       & color     \\ \hline
 %\citep{Gutub2021_3}                             & DWT                                                                                                                     & binary data       & color     \\ \hline

\end{tabular}
%}
\end{table}

As mentioned above, images are of three kinds; binary, grayscale, and color. A grayscale image has more information than a binary image. Similarly, a color image has more information than a grayscale image. Thus, hiding a color secret image is more challenging than hiding a grayscale secret image, which is more challenging than hiding a binary secret image. Similarly, applying this concept to the cover image, we see a reverse sequence; see Table \ref{tbl:imageType}. We focus on the middle case here, i.e., when both the secret images and the cover image are grayscale, which is considered challenging. {\color{myrosewood}This forms our \emph{third} focus.}

\begin{table}[!h]
	\centering
	%\footnotesize
	\caption{Image types and levels of challenge.}
	\label{tbl:imageType}
	\setlength{\tabcolsep}{2.5pt}
	\begin{tabular}{*3{c@{\qquad}}c}
	\toprule
	\textbf{Image Type} & \textbf{More Challenging} & \textbf{Medium Challenging} & \textbf{Less Challenging} \\ \midrule
	Secret Image        & Color                     & Grayscale                   & Binary                    \\ %\midrule
	Cover Image         & Binary                    & Grayscale                   & Color                     \\ \bottomrule
	\end{tabular}
\end{table}

The difficulty in designing a good steganography scheme for hiding secret images into a cover image is increasing the embedding capacity of the scheme while preserving the quality of the resultant stego-image and extracted secret images as well as making the scheme resistant to steganographic attacks. 
%Usually, the more the number of secret images to be embedded (which often translates to heavier secret images), the lower the quality of the obtained stego-image. 
Hence, we need to balance these competing requirements. 
%{\color{myrosewood}Until now, in most works, researchers have embedded up to two secret images in a cover image when employing the transform domain based approach, which is our focus. Using the same transform domain based approach, no one has looked up at embedding three or four secret images.} 
Here, not just the number of secret images but the total size of the secret images is also important. To capture this requirement of number as well as size, a metric of bits per pixel (bpp) is used.

In this work, we present a novel image steganography scheme wherein up to four images can be hidden in a single cover image. The size of the length and the width of a secret image is about half of the length and the width of the cover image, respectively, which results in a very high \si{bpp} capacity. No one has attempted hiding up to four secret images in a cover image {\color{myrosewood}with the transform domain based approach} until now, and those who have attempted hiding one, or two images have also not achieved the level of embedding capacity that we do. While enhancing the capacity as discussed above, the quality of our stego-image does not deteriorate much. Also, we do not need any cover image data to extract secret images on the receiver side, which is commonly required with other schemes. We do require some algorithmic settings on the receiver side, however, these can be communicated to the receiver separately. Thus, this makes our scheme more secure.
%\mcs{But we must know the matrix $\Phi$ and the algorithmic constants!}

{\color{myyaleblue}Let us consider the example of telediagnosis that refers to remote diagnosis. In this, medical images are distributed to some doctors for analyses and recommendations. During distribution, an unauthorized person can access these images and misuse them. To make this distribution process more secure, instead of directly sharing images, these can be hidden in a cover image using our steganography scheme and then the obtained stego-image can be shared. In this example, multiple secret images need to be shared (we consider sharing a maximum of four medical images). The existing transform based steganography schemes, which are inherently resistant to steganographic attacks, do not have such an embedding capacity. If we try to increase their capacity, then the quality of stego-image or extracted secret images gets degraded.}

{\color{myviolet}The most novel feature of our innovative scheme is that it is a combination of different components that helps us achieve the competing goals of increasing embedding capacity, good quality stego-image as well as extracted secret images, and high resistance to steganographics attacks. Each of these components is discussed next.}

The \textit{first} component, i.e., hiding of secret images, consists of the parts below. \\
(\romannum{1}) We perform sub-sampling on a cover image to obtain four sub-images of the cover image. \\
(\romannum{2}) We perform block-wise sparsification of each of these four sub-images using DCT (Discrete Cosine Transform) and form respective vectors. \\
(\romannum{3}) We represent each vector in two groups based upon large and small coefficients, and then {\color{myyaleblue}oversample each of the resultant (or generated) sparse vector using a measurement matrix based linear measurements.} The oversampling at this stage leads to sparse approximation. \\
(\romannum{4}) We repeat the second step above for each of the secret images. \\
(\romannum{5}) We embed DCT coefficients from the four secret images into ``a set'' of linear measurements obtained from the four sub-images of the cover image using our new embedding rule. 

{Amongst these parts, (\romannum{1})--(\romannum{2}) have been used in \citep{Steg_Rohit, Liu, Pan} while (\romannum{3})--(\romannum{5}) are new.}
% Sparse approximation along with the our novel embedding rule gives \textit{higher embedding capacity} to our scheme and add a layer of \textit{security} to it as well.}

\textit{Second}, we generate the stego-image from these modified measurements by using the Alternating Direction Method of Multipliers (ADMM) to solve the Least Absolute Shrinkage and Selection Operator (LASSO) formulation of the underlying minimization problem. This method has a fast convergence, is easy to implement, and also is extensively used in image processing. Here, the optimization problem is an $\ell_1$-norm minimization problem, and the constraints comprise an \textit{over-determined system of equations} \citep{SparseApproximation_Overdetermined}. {\color{myviolet}Use of this component in steganography is first of its kind as well.}

\textit{Third}, we extract the secret images from the stego-image using our proposed extraction rule, which is {the} reverse of our embedding rule mentioned above. As mentioned earlier, we do not require any information about the cover image while doing this extraction, which makes the process blind. {\color{myviolet}Since our embedding procedure, as mentioned above, is new, thus the extraction part is also new.} We call our scheme SABMIS (Sparse Approximation Blind Multi-Image Steganography), and is described in Section \ref{Sec:L1SABMIS_Proposed}.

For performance evaluation, in Section \ref{Sec:L1SABMIS_NumericalExp} we perform extensive experiments on a set of standard images. We {\it first} compute the embedding capacity of our scheme, which turns out to be very good. {\it Next}, we check the quality of the stego-images by comparing them with their corresponding cover images. We use both a visual measure and a set of numerical measures for this comparison. The numerical measures used are: Peak Signal-to-Noise Ratio (PSNR) value, Mean Structural Similarity (MSSIM) index, Normalized Cross-Correlation (NCC) coefficient, entropy, and Normalized Absolute Error (NAE). The results show very little deterioration in the quality of the stego-images. 

{\it Further}, we visually demonstrate the high quality of the extracted secret images by comparing them with the corresponding original secret images. {\it Also}, via experiments, we support our conjecture that our scheme is resistant to steganographic attacks. \textit{Next}, we demonstrate efficiency of our scheme by providing timing data. \textit{Finally}, we present application of our scheme on real-life data in-turn demonstrating its usefulness.

Also, we exhaustively compare SABMIS with competing schemes to demonstrate that it is among the best. For the sake of better exposition, this comparison is given in Introduction itself (see subsection below). Finally, in Section \ref{Conclusion}, we discuss conclusions and future work.

%{\it Finally}, we compare the embedding capacity of our scheme and the quality of our stego-images with the corresponding data from competing schemes available in the literature. The quality of the extracted secret images and experimentation for resistance to steganographic attacks are not common in existing works, and hence, we are unable to perform these two comparisons. The superiority of our scheme over past works is summarized below.
% in four paragraphs where one, two, three, and four secret images are embedded in a cover image, respectively.

%%%%%%%%%%%%%%%%% Comparison with existing methods %%%%%%%
\par
\subsection{Comparison with Past Work}\label{Performance Comparison SABMIS}
Here, we predominately compare our SABMIS scheme with the existing steganography schemes for the embedding capacity, the quality of stego-images, and resistant to steganographic attacks. For the stego-image quality comparison, since most works have computed PSNR values only, we use only this metric for our analysis. Although we check the quality of the extracted secret images by comparing them with the corresponding original secret images (as earlier), this check is not common in the existing works. Hence, we do not perform this comparison.

In the literature, there exist multiple transform-based steganography schemes that hide one or two secret images. Hence, in Table \ref{Table:Performance comparison with various other transform steganography schemes SABMIS} we compare our SABMIS scheme using the above mentioned metrics with such competing schemes. Recall, that like our SABMIS scheme these schemes are inherently resistant to steganographic attacks as well. 

{\color{myrosewood}As evident from Table \ref{Table:Performance comparison with various other transform steganography schemes SABMIS}, for the case of hiding one secret image, we compare with the best work of this category \citep{S_Arunkumar_MedicalImage_DWT_2019}. Here, as for us, by using a transform based approach, a grayscale secret image is hidden into a grayscale cover image. The authors in \citep{S_Arunkumar_MedicalImage_DWT_2019} and our scheme both achieve an embedding capacity of \SI{2}{bpp}. When comparing the stego-image and the corresponding cover image, \citep{S_Arunkumar_MedicalImage_DWT_2019} achieve a PSNR value of \SI{49.69}{dB} (when experimented with eight cover images) while we achieve a lower PSNR value of \SI{41.64}{dB} (when experimenting with a higher number of cover images, i.e., ten). PSNR values over \SI{30}{dB} are considered good \citep{Ste_2020, Zhang2013, Liu}. Although, the scheme by \citep{S_Arunkumar_MedicalImage_DWT_2019} is superior than ours for hiding one secret image, it does not scale for the case of hiding multiple secret images, which we do (please see below).} 

For the case of hiding two secret images, we again compare with the best work of this category \citep{S_Hemalatha}. Here, using the transform based approach, two grayscale secret images are hidden into a color cover image. This setup is easier than our case where using a transform based approach, we embed two grayscale secret images into a grayscale cover image (see Table \ref{tbl:imageType}). The authors in \citep{S_Hemalatha} achieve an embedding capacity of \SI{1.33}{bpp} while we achieve a higher embedding capacity of \SI{4}{bpp}. When comparing the stego-image and the corresponding cover image, \citep{S_Hemalatha} achieve a PSNR value of \SI{44.75}{dB} (when experimented with only two cover images) while we achieve a lower PSNR value of \SI{38.74}{dB} (when experimenting with a higher number of cover images, i.e., ten). 
{\color{myrosewood}To sum-up, our scheme is better than the one by \citep{S_Hemalatha} because of the below reasons. \\
In-terms of the quality of the scheme,
\vspace*{-0.3cm}
\begin{enumerate}[label=\alph*)]  %[topsep=0pt,itemsep=-1ex,partopsep=1ex,parsep=1ex][label=\alph*)]
\itemsep0em 
\item we target a harder problem than \citep{S_Hemalatha}, and
\item we achieve a higher embedding capacity than \citep{S_Hemalatha}.
\end{enumerate}
\vspace*{-0.3cm}
In-terms of the validation of the scheme,
\vspace*{-0.3cm}
\begin{enumerate}[label=\alph*)]  %[topsep=0pt,itemsep=-1ex,partopsep=1ex,parsep=1ex][label=\alph*)]
\itemsep0em 
\item we experiment with a large number of cover images (ten as compared to two in \citep{S_Hemalatha}),
\item as discussed earlier, we obtain PSNR values over 30 dB of stego-images, which are considered acceptable, and
\item we check the quality of stego-image on greater number of numerical measures (five as compared to one in \citep{S_Hemalatha}).
\end{enumerate}
}

When using the transform-based approach, no one has hidden three or four secret images in a cover image. To demonstrate the broad applicability of our scheme, in Table \ref{Table:Performance comparison with various other LSB steganography schemes SABMIS}, we compare our SABMIS scheme using the above discussed metrics with the best spatial domain-based scheme that hide three and four secret images. Recall that, unlike our SABMIS scheme, these schemes are not intrinsically resistant to steganographic attacks. Please note that in the current scenario of transmitting stego-data over the internet, security is of paramount importance. 

{\color{myrosewood}As evident from Table \ref{Table:Performance comparison with various other LSB steganography schemes SABMIS}, 
for the case of hiding three secret images, we compare with the best work of this category \citep{Prashanti}. Here, three binary secret images are hidden into a grayscale cover image. As for the above case, this setup is easier than our case of hiding three grayscale secret images into a grayscale cover image (again see Table \ref{tbl:imageType}). The authors in \citep{Prashanti} achieve an embedding capacity of \SI{1}{bpp} while we achieve a higher embedding capacity of \SI{6}{bpp}. When comparing the stego-image and the corresponding cover image, \citep{Prashanti} achieve a PSNR value of \SI{46.36}{dB} (when experimented with only two cover images) while we achieve a lower PSNR value of \SI{37.17}{dB} (when experimenting with a higher number of cover images, i.e., ten). To sum-up, our scheme is superior than the one by \citep{Prashanti} because of the below reasons. \\
In-terms of the quality of the scheme, 
\vspace*{-0.3cm}
\begin{enumerate}[label=\alph*)]  %[topsep=0pt,itemsep=-1ex,partopsep=1ex,parsep=1ex][label=\alph*)]
\itemsep0em 
\item we target a harder problem than \citep{Prashanti}, 
\item we achieve a higher embedding capacity than \citep{Prashanti}, and 
\item we further improve the security of the inherently steganographic attack resistant transform based schemes.
\end{enumerate}
\vspace*{-0.3cm}
In-terms of the validation of the scheme,
\vspace*{-0.3cm}
\begin{enumerate}[label=\alph*)]  %[topsep=0pt,itemsep=-1ex,partopsep=1ex,parsep=1ex][label=\alph*)]
\itemsep0em 
\item we experiment with a large number of cover images (ten as compared to two in \citep{Prashanti}), 
\item as discussed earlier, we obtain PSNR values over 30 dB of stego-images, which are considered acceptable, 
\item we check the quality of stego-image on greater number of numerical measures (five as compared to one in \citep{Prashanti}), 
\item and we demonstrate the good quality of extracted secret images, which \citep{Prashanti} do not.
\end{enumerate}

%\vspace*{-0.3cm}
%\begin{itemize}
%\item in-term of quality
%\vspace*{-0.3cm}
%
%\begin{enumerate}[label=\alph*)]  %[topsep=0pt,itemsep=-1ex,partopsep=1ex,parsep=1ex][label=\alph*)]
%\itemsep0em 
%\item we are targeting a harder problem than \citep{Prashanti},
%\item we achieve a higher embedding capacity than \citep{Prashanti}, and
%\item {\color{myred}our scheme improves the inherently resistant to steganographic attacks transform based schemes.}
%\end{enumerate}
%
%\vspace*{-0.3cm}
%\item in-term of validity
%\vspace*{-0.3cm}
%
%\begin{enumerate}[label=\alph*)]  %[topsep=0pt,itemsep=-1ex,partopsep=1ex,parsep=1ex][label=\alph*)]
%\itemsep0em 
%\item we have experimented with a large number of cover images (ten as compared to two in \citep{Prashanti}),
%\item as discussed earlier, we obtain PSNR values over 30 dB of stego-images, which are considered acceptable,
%\item we check the quality of stego-image on more number of numerical measures as compared to \citep{Prashanti}, and
%\item we demonstrate the good quality of extracted secret images, which \citep{Prashanti} do not.
%%our scheme is inherently as well as designed to be resistant to steganographic attacks.
%\end{enumerate}
%\end{itemize}
}

{\color{myrosewood}Next, we compare with the best scheme that hide four secret images in a cover image, i.e., \citep{Ste_Ios_2006}. As for our case, all images (secret and cover) are grayscale. The authors in \citep{Ste_Ios_2006} achieve an embedding capacity of \SI{12}{bpp} while we achieve a lower embedding capacity of \SI{8}{bpp}. When comparing the stego-image and the corresponding cover image, \citep{Ste_Ios_2006} achieve a PSNR value of \SI{34.80}{dB} (when experimented with five cover images) while we achieve a higher PSNR value of \SI{35.66}{dB} (when experimenting with a higher number of cover images, i.e., ten). To sum-up, our scheme is better than the one by \citep{Ste_Ios_2006} because of the below reasons. \\
In-terms of the quality of the scheme, 
\vspace*{-0.3cm}
\begin{enumerate}[label=\alph*)]  %[topsep=0pt,itemsep=-1ex,partopsep=1ex,parsep=1ex][label=\alph*)]
\itemsep0em 
\item our embedding capacity, although lower than \citep{Ste_Ios_2006}, is on the higher side, 
\item we obtain higher PSNR values of stego-images as compared to those in \citep{Ste_Ios_2006}, 
\item and we further improve the security of the inherently steganographic attack resistant transform based schemes.
\end{enumerate}
\vspace*{-0.3cm}
In-terms of the validation of the scheme, 
\vspace*{-0.3cm}
\begin{enumerate}[label=\alph*)]  %[topsep=0pt,itemsep=-1ex,partopsep=1ex,parsep=1ex][label=\alph*)]
\itemsep0em 
\item we experiment with a large number of cover images (ten as compared to five in \citep{Ste_Ios_2006}), 
\item we check the quality of stego-image on greater number of numerical measures (five as compared to one in \citep{Ste_Ios_2006}), 
\item and we demonstrate the good quality of extracted secret images, which \citep{Ste_Ios_2006} do not.
\end{enumerate}

%\vspace*{-0.3cm}
%\begin{itemize}
%\item in-term of quality
%\vspace*{-0.3cm}
%\begin{enumerate}[label=\alph*)]  %[topsep=0pt,itemsep=-1ex,partopsep=1ex,parsep=1ex][label=\alph*)]
%\itemsep0em 
%\item our embedding capacity, although lower than \citep{Ste_Ios_2006}, is on the higher side, 
%\item we obtain higher PSNR values of stego-images as compared to those in \citep{Ste_Ios_2006}, and
%\item {\color{myred}our scheme improves the inherently resistant to steganographic attacks transform based schemes.}

%
%\vspace*{-0.3cm}
%\item in-term of validity
%\vspace*{-0.3cm}
%
%\begin{enumerate}[label=\alph*)]  %[topsep=0pt,itemsep=-1ex,partopsep=1ex,parsep=1ex][label=\alph*)]
%\itemsep0em 
%\item we have experimented with a large number of cover images (ten as compared to five in \citep{Ste_Ios_2006}),
%\item we check the quality of stego-image on more number of numerical measures as compared to \citep{Ste_Ios_2006}, and
%\item we demonstrate the good quality of extracted secret images, which \citep{Ste_Ios_2006} do not.
%%our scheme is inherently as well as designed to be resistant to steganographic attacks.
%\end{enumerate}
%\end{itemize}
}

\begin{table}[!h]
\centering
\footnotesize
\caption{Performance comparison of our SABMIS scheme with competing transform-based steganography schemes, which are inherently resistant to steganographic attacks.}
\label{Table:Performance comparison with various other transform steganography schemes SABMIS}
\setlength{\tabcolsep}{2.5pt}
\begin{tabular}{|c|c|c|c|c|c|c|c|}
\hline
\textbf{\begin{tabular}[c]{@{}c@{}}No. \\ of \\ secret \\ images\end{tabular}} & \textbf{\begin{tabular}[c]{@{}c@{}}Steganography \\Scheme\end{tabular}}         & \textbf{\begin{tabular}[c]{@{}c@{}}Type of \\ secret \\ image\end{tabular}} & \textbf{\begin{tabular}[c]{@{}c@{}}Type of \\ cover \\images\end{tabular}} & \textbf{\begin{tabular}[c]{@{}c@{}}EC \\ (in \si{bpp})\end{tabular}} & \textbf{\begin{tabular}[c]{@{}c@{}}(Avg. PSNR, \\ No. of \\ Cover\\ Images)\end{tabular}} & \textbf{\begin{tabular}[c]{@{}c@{}}Max. \\PSNR\end{tabular}} & \textbf{\begin{tabular}[c]{@{}c@{}}Resistant \\ to stegan-\\ ographic\\ attacks?\end{tabular}} \\ \hline
\multirow{2}{*}{1}                                                                 & \citep{S_Arunkumar_MedicalImage_DWT_2019} & {\color{myrosewood}Grayscale}                                                               & Grayscale                                                                    & {\color{myrosewood}2}                                                                              & (49.69, 8)            & 50.15                                               & Yes                                                                                            \\ \cline{2-8}
                                                                                   & {SABMIS}                       & Grayscale                                                               & Grayscale                                                                 & 2                                                                                  & (41.64, 10)        & 46.25                                       & Yes                                                                                            \\
        \Xhline{3\arrayrulewidth}%\hline
\multirow{2}{*}{2}                                                                 & \citep{S_Hemalatha}                          & Grayscale                                                                   & Color                                                                 & 1.33                                                                               & (44.75, 2)  & 44.80                                                          & Yes                                                                                            \\ \cline{2-8}
                                                                                   & {SABMIS}                       & Grayscale                                                               & Grayscale                                                                 & 4                                                                                  & (38.74, 10)         & 42.60                                                 & Yes                                                                                            \\
         \Xhline{3\arrayrulewidth}%\hline
         
\end{tabular}
\end{table}

\begin{table}[!h]
\centering
\footnotesize
\caption{Performance comparison of our SABMIS scheme with competing spatial domain-based steganography schemes, which are not inherently resistant to steganographic attacks.}
\label{Table:Performance comparison with various other LSB steganography schemes SABMIS}
\setlength{\tabcolsep}{2.5pt}
\begin{tabular}{|c|c|c|c|c|c|c|c|}
\hline
\textbf{\begin{tabular}[c]{@{}c@{}}No. \\ of \\ secret \\ images\end{tabular}} & \textbf{\begin{tabular}[c]{@{}c@{}}Steganography \\Scheme\end{tabular}}         & \textbf{\begin{tabular}[c]{@{}c@{}}Type of \\ secret \\ image\end{tabular}} & \textbf{\begin{tabular}[c]{@{}c@{}}Type of \\ cover \\images\end{tabular}} & \textbf{\begin{tabular}[c]{@{}c@{}}EC \\ (in \si{bpp})\end{tabular}} & \textbf{\begin{tabular}[c]{@{}c@{}}(Avg. PSNR, \\ No. of \\ Cover\\ Images)\end{tabular}} & \textbf{\begin{tabular}[c]{@{}c@{}}Max. \\PSNR\end{tabular}} & \textbf{\begin{tabular}[c]{@{}c@{}}Resistant \\ to stegan-\\ ographic\\ attacks?\end{tabular}} \\ \hline
\multirow{2}{*}{3}                                                                 & \citep{Prashanti}                          & Binary                                                               & Grayscale                                                                     & {\color{myrosewood}1}                                                                                  & (46.36, 2)        & 46.38                                                   & No                                                                                             \\ \cline{2-8}
                                                                                   & {SABMIS}                       & Grayscale                                                               & Grayscale                                                                 & 6                                                                                  & (37.17, 10)           & 41.06                                                 & Yes         
                                                                                                                                                                      \\
        \Xhline{3\arrayrulewidth}%\hline
\multirow{2}{*}{4}                                                                 & {\color{myrosewood}\citep{Ste_Ios_2006}}                           & {\color{myrosewood}Grayscale}                                                                   & {\color{myrosewood}Grayscale}                                                                     & {\color{myrosewood}12}                                                                                  & {\color{myrosewood}(34.80, 5)}       & {\color{myrosewood}34.82}                                                   & {\color{myrosewood}No}                                                                                             \\ \cline{2-8}
                                                                                   & {SABMIS}                       & Grayscale                                                               & Grayscale                                                                 & 8                                                                                  & (35.66, 10)      & 39.74                                                    & Yes         \\ \hline
\end{tabular}
\end{table}

\section{Proposed Approach}\label{Sec:L1SABMIS_Proposed}
%%%%%% First part %%%%%%%%
Our sparse approximation based blind multi-image steganography scheme consists of the following components: (\romannum{1}) Hiding of secret images leading to the generation of the stego-data. (\romannum{2}) Construction of the stego-image. (\romannum{3}) Extraction of secret images from the stego-image. These parts are discussed in the respective subsections below.

\subsection{Hiding Secret Images}\label{Subsec:L1SABMIS_SecretImagesEmbedding}
\begin{figure}[!h]
	\centering
		\includegraphics[width=0.99\textwidth]{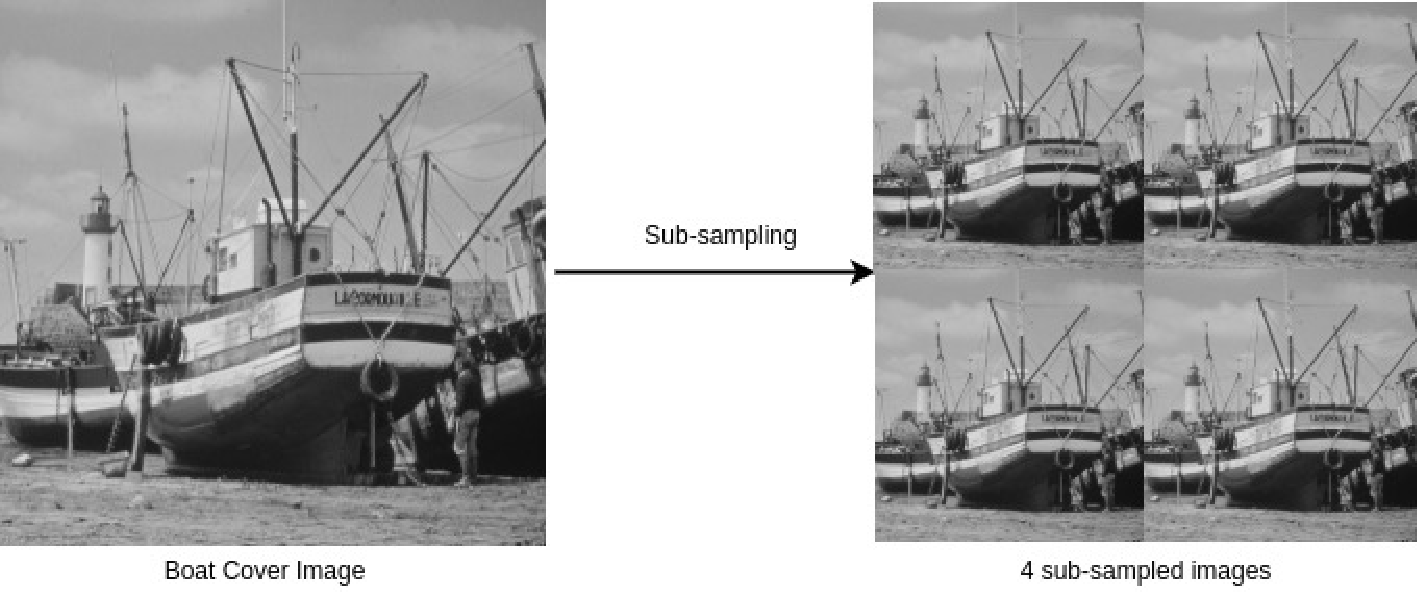}
		\caption{A cover image and its 4 sub-sampled images \citep{standard-test-images-for-Image-Processing} \url{https://github.com/mohammadimtiazz/standard-test-images-for-Image-Processing}.}
		\label{fig:subsampled images}
\end{figure}

First, we perform sub-sampling of the cover image to obtain four sub-images. {\color{myblue}This type of sampling is done because we are hiding up to four secret images.} Let $CI$ be the cover image of size $r\times r$. Then, the four sub-images each of size $\frac{r}{2} \times \frac{r}{2}$ are obtained as follows \citep{Pan}:
%\begin{equation}
\begin{linenomath*}
\begin{align}\label{eq:sampling}
& {CI}^1(n_1,n_2) = CI(2n_1-1,2n_2-1),   \refstepcounter{equation} \subeqn \\ %\nonumber
& {CI}^2(n_1,n_2) = CI(2n_1,2n_2-1),   \subeqn \\
& {CI}^3(n_1,n_2) = CI(2n_1-1,2n_2),   \subeqn \\
& {CI}^4(n_1,n_2) = CI(2n_1,2n_2),   \subeqn \label{eq:sampling_eq4}
\end{align}
\end{linenomath*}
%\end{equation}
where $CI^k$, $\text{for} \; k=\{1, 2, 3, 4\}$, are the four sub-images; $n_1,n_2 = 1,2,\hdots, \frac{r}{2}$ (in our case, $r$ is divisible by~$2$); and $CI({}\cdot{},{}\cdot{})$ is the pixel value at $({}\cdot{},{}\cdot{})$. A cover image and the corresponding four sub-sampled images are shown in Figure \ref{fig:subsampled images}.
\par

\begin{figure}[!h]
	\centering
		\includegraphics[width=0.49\textwidth]{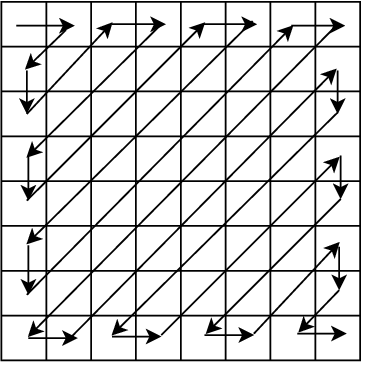}
		\caption{Zig-zag scanning order for a block of size $8 \times 8$.}
		\label{fig:zigzag scanning order}
\end{figure}

Originally, these sub-images are not sparse; hence, next, we perform block-wise sparsification of each of these images. For this, we divide each sub-image into blocks of size $b\times b$ and obtain $\frac{r^2}{4\times b^2}$ blocks for each sub-image (in our case, $b$ divides $r$). %Now, we sparsify each block using the discrete cosine transformation. That is,
Now, we apply discrete cosine transformation to each block. That is,
\begin{linenomath*}
\begin{equation}\label{eq:propose sparsification SABMIS}
%s_i = \Psi^T x_i,
s_i = DCT(x_i),
\end{equation}
\end{linenomath*}
where $i=1,2,\hdots,\frac{r^2}{4\times b^2}$, $x_i$ and $s_i$ are the $i^{th}$ original and sparse blocks of the same size, i.e, $b\times b$, respectively, and DCT is the Discrete Cosine Transform. Further, we pick the final sparse blocks using a zig-zag scanning order as used in our earlier work \citep{Steg_Rohit}, and obtain corresponding sparse vectors each of size $b^2 \times 1$. The zig-zag scanning order for a block of size $8 \times 8$ is shown in Figure \ref{fig:zigzag scanning order}. This order helps us to arrange the DCT coefficients with the set of large coefficients first, followed by the set of small coefficients, {\color{myblue}which assists in the preservation of a good quality stego-image}.

Next, we represent each vector in two groups based upon large (say $\#p_1$\label{p1_page_ref_section}) and small (say $\#p_2$) coefficients, i.e., $s_{i,u}\in \mathbb{R}^{p_1}$ and $s_{i,v}\in \mathbb{R}^{p_2}$, where $p_1 \leq p_2$. Each of these vectors is sparse and $p_1 + p_2 = b^2$. {\color{myyaleblue}Further, we oversample each sparse vector using linear measurements as below.}
\begin{linenomath*}
\begin{equation}\label{eq: propose measurements SABMIS}
y_i=\begin{bmatrix}y_{i,u}\\ y_{i,v} \end{bmatrix}=
\begin{bmatrix}
s_{i,u}\\ \Phi s_{i,v}
\end{bmatrix},
\end{equation}
\end{linenomath*}
where $y_i\in \mathbb{R}^{({p_1+p_3})\times 1}$ is the set of linear measurements, and $\Phi \in \mathbb{R}^{p_3\times p_2}$ is the column normalised measurement matrix consisting of normally distributed random numbers with $p_3>p_2$ and $p_3\in \mathbb{N}$ (i.e., the sparse coefficients are oversampled) \footnote{\label{note1}In the experimental results section, we show how to experimentally pick these coefficients.}. 
This oversampling helps us to perform sparse approximation. By employing this approximation (along with our novel embedding rule discussed towards the end of this subsection), we achieve a higher embedding capacity. Moreover, our approach gains an extra layer of security because the linear measurements include measurement-matrix encoded small coefficients of the sparse vectors obtained after DCT. Since the distribution of coefficients of the generated sparse vectors is almost the same for all the blocks of an image, we use the same measurement matrix for all the blocks.

\par

Next, we perform processing of the secret images for hiding them into the cover image. Let the size of each secret image be $m \times m$. Initially, we perform block-wise DCT of each of these images and obtain their corresponding DCT coefficients. Here, the size of each block taken is $l\times l$, and hence, we have $\frac{m^2}{l^2}$ blocks for each secret image. In our case, $l$ divides $m$, and we ensure that $\frac{m^2}{l^2}$ will be less than or equal to $\frac{r^2}{4\times b^2}$\label{page_coeffm2/l2} so that the number of blocks of the secret image is less than or equal to the number of blocks of a cover sub-image. Thereafter, we arrange these DCT coefficients as a vector in the earlier discussed zig-zag scanning order. Let $t_{\hati} \in R^{l^2\times 1}$, for $\hati=1,2,\hdots,\frac{m^2}{l^2}$, be the vector representation of the DCT coefficients of one secret image. We pick the initial $p_4$ \label{p4_page_ref_section} DCT coefficients with relatively larger values (out of the available $l^2$ coefficients) for hiding\footref{note1}, where $p_4\in \mathbb{N}$. {\color{myyaleblue}Omitting the remaining coefficients ($l^2 - p_4$) does not significantly deteriorate the quality of the extracted secret image.}

Here, we show {the} hiding of only one secret image into one sub-image of the cover image. However, in our steganography scheme, we can hide a maximum of four secret images, one in each of the four sub-images of the cover image, which is demonstrated in the experimental results section. If we want to hide less than four secret images, we can randomly select the corresponding sub-images from the available four.

Next, using our novel embedding rule (discussed below), we hide the chosen $p_4$ DCT coefficients of the secret image into a selected set of $p_1 + p_3$ linear measurements obtained from the sub-image of the cover image, leading to the generation of the stego-data (we ensure that $p_4$ is less than $p_1 + p_3$). 
%The selected linear measurements (discussed in Table \ref{Table:detail of embedding coefficients into the linear measurement}) are chosen to give the best results, i.e., less deterioration in the quality of the stego-image plus the extracted secret image and more security.

\begin{table}[!h]
\centering
\footnotesize
\caption{The detail of hiding secret image coefficients into the linear measurement coefficients of the cover image.}
\label{Table:detail of embedding coefficients into the linear measurement}
{\color{myyaleblue}
\begin{tabular}{|ccc|}
\hline
\multicolumn{3}{|c|}{Secret Image Coefficient Indices}                              \\ \hline
\multicolumn{1}{|c|}{1} & \multicolumn{1}{c|}{2 to c} & $c+1$ to $p_4$ \\ \hline
\multicolumn{3}{|c|}{Companion Linear Measurement Coefficient Indices}                            \\ \hline
\multicolumn{1}{|c|}{$p_1 - 2c$} & \multicolumn{1}{c|}{$p_1 -  2c + 1$ to $p_1 - c  - 1$} & $p_1 + c + 1$ to $p_1 + p_4 $ \\ \hline
\multicolumn{3}{|c|}{Replaced Linear Measurement Coefficient Indices}                            \\ \hline
\multicolumn{1}{|c|}{$p_1$} & \multicolumn{1}{c|}{$p_1 -  c + 1$ to $p_1 - 1$} & $p_1 + p_4 + 1$ to $p_1 + 2\times p_4 - c$  \\ \hline
\end{tabular}
}
\end{table}

\begin{figure}[]
	\centering
		\includegraphics[width=\textwidth]{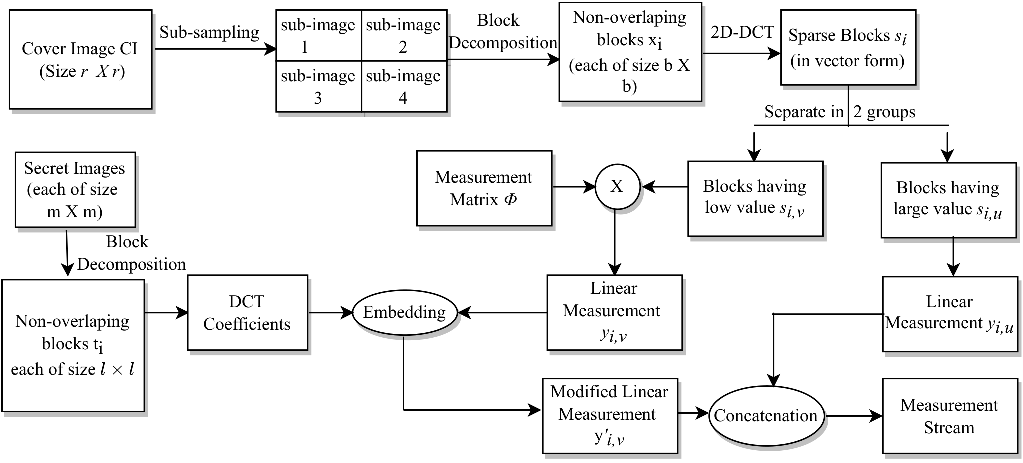}
		\caption{The embedding process.}
		\label{Fig:The embedding process}
\end{figure}

{\color{myyaleblue}
%For enhanced security, we hide secret image data into the original cover image linear measurements resulting in the modified linear measurements, also called the stego-data, in three parts. These three groups of secret image coefficients are listed in Table \ref{Table:detail of embedding coefficients into the linear measurement}. 
We hide secret image data into the cover image by taking linear combinations of each secret image coefficient with a companion linear measurement coefficient of the cover image. These linear combinations replace certain other linear coefficients of the cover image to obtain the so called stego-data (subsequently, stego-image). The three groups of index coefficients are listed in Table \ref{Table:detail of embedding coefficients into the linear measurement}. 

%The choice of where to embed the secret image coefficients into the original linear measurements is based so as to achieve high quality of stego-image and extracted secret image. Recall that the original linear measurements now consist of $p_1 + p_3$ coefficients where we can hide secret image coefficients. Out of these, the initial coefficients are of large value; hence hiding there would lead to deterioration in both the quality of the stego-image and the extracted secret image. Hence, we leave those initial sets of original linear measurement coefficients and start first hiding at the index $p_1 -  2c$, which is again evident in Table \ref{Table:detail of embedding coefficients into the linear measurement}. The rest of the secret image coefficients are hidden into the original linear measurement coefficients in a staggered manner. This helps us achieve a higher level of security. Finally, if we look at the modified linear measurement coefficients listed in Table \ref{Table:detail of embedding coefficients into the linear measurement} , these are slightly different from the original linear measurement coefficients mentioned in the same table. The reason for this is that we want our extraction rule (discussed in section \ref{Subsec:L1SABMIS_SecretImagesExtraction}) to be as less lossy as possible. 
The data in Table \ref{Table:detail of embedding coefficients into the linear measurement} is based upon three design choices as below. 
\vspace{-0.3cm}
\begin{enumerate}[label=\alph*)]  %[topsep=0pt,itemsep=-1ex,partopsep=1ex,parsep=1ex][label=\alph*)]
\itemsep0em 
\item As can be seen from Table \ref{Table:detail of embedding coefficients into the linear measurement}, we divide each group of coefficients into three ranges in a staggered manner to achieve a higher level of security.

\item The specific choice of indices in the second and fourth rows of Table \ref{Table:detail of embedding coefficients into the linear measurement} is made so as to hide secret image coefficients in relatively small valued cover image coefficients (companion linear measurement coefficients). This results in relatively improved quality stego-image.  

\item In Table \ref{Table:detail of embedding coefficients into the linear measurement}, the replaced linear measurement coefficient indices differ just slightly from the chosen companion coefficient indices (fourth and sixth rows respectively). The reason for this is that we want our extraction rule (discussed in section \ref{Subsec:L1SABMIS_SecretImagesExtraction}) to be as less lossy as possible, resulting in less deteriorated extracted secret images.
\end{enumerate}

The whole process is given in \textbf{Algorithm \ref{alg:Embedding rule SABMIS}}. Specifically, the indices discussed in Table \ref{Table:detail of embedding coefficients into the linear measurement} are given on line \ref{alg:eq 1 embedding rule},  lines \ref{alg:eq 2 embedding rule} -- \ref{alg:eq2 embedding rule end}, and lines \ref{alg:eq 3 embedding rule} -- \ref{alg:eq 3 embedding rule end} of this algorithm, respectively. The block diagram for this complete data embedding process is given in Figure \ref{Fig:The embedding process}. A small numerical example, which further explains this hiding process is given in Appendix \ref{appendix:A small numerical example of working of our embedding process}.
%Recall that $p_1 + p_3$ linear measurement coefficients are available for hiding secret images. Out of these, the initial coefficients are of large value. To avoid deterioration in both the quality of the stego-image and the extracted secret images, we keep those initial coefficients and start replacing at index $p_1 - c + 1$ (second range). 
}

{
	\begin{algorithm*}[!t]
	%\scriptsize %\small, \footnotesize, \scriptsize, or \tiny
	%\footnotesize
		\caption{Embedding Rule}\label{alg:Embedding rule SABMIS}
		\begin{algorithmic}[1]
			\renewcommand{\algorithmicrequire}{\textbf{Input:}}
			\renewcommand{\algorithmicensure}{\textbf{Output:}}
			\Require \quad
			\begin{itemize}
				\item $y_i$: Sequence of linear measurements of the cover image with $i=1,2,\hdots,\frac{r^2}{4\times b^2}$.
				\item $t_{\hati}$: Sequence of transform coefficients of the secret image with $\hati=1,2,\hdots,\frac{m^2}{l^2}$.
				\item The choice of our $r$, $b$, $m$, and $l$ is such that $\frac{m^2}{l^2}$ is less than or equal to $\frac{r^2}{4\times b^2}$.
				\item $p_1$ and $p_4$ are lengths of certain vectors defined on pages \pageref{p1_page_ref_section} and \pageref{p4_page_ref_section}, respectively.
				\item $\alpha$, $\beta$, $\gamma$, and $c$ are algorithmic constants that are chosen based upon experience. The choices of these constants are discussed in the experimental results sections.
			\end{itemize}

			\Ensure \quad
			\begin{itemize}
				\item $y'_i$: The modified version of the linear measurements with $i=1,2,\hdots,\frac{r^2}{4\times b^2}$.
			\end{itemize}
		%	\\ \textit{Initialisation}:
		    \State Initialize ${y'_i}$ to $y_i$, where $i=1,2,\hdots,\frac{r^2}{4\times b^2}$.
			\For {${\hati} = 1$ to $\frac{m^2}{l^2}$}
			\vspace{0.1cm}
                \State // Embedding of the first coefficient.\label{alg:eq 1 embedding rule}
                \vspace{0.1cm}
		        	%\begin{align}
                 \Statex  \qquad \hspace{-0.3cm} $y'_{\hati}(p_1) = y_{\hati}(p_1-2c) + \alpha \times t_{\hati}(1)$.  %\nonumber
                 \vspace{0.1cm}
                  %\end{align}

                \For {$j = p_1 - c + 1$ to $p_1 - 1$} \label{alg:eq 2 embedding rule}
                \vspace{0.1cm}
				\State // Embedding of the next $c-1$ coefficients.
				\vspace{0.1cm}
                    % \begin{align}
                   \Statex \qquad \quad $y'_{\hati}(j) = y_{\hati}(j-c) + \beta \times t_{\hati}(j-p_1+c+1)$. % \nonumber
                  %\end{align}
                  \vspace{0.1cm}
                \EndFor												\label{alg:eq2 embedding rule end}
                \vspace{0.1cm}
                \For {$k = p_1 + p_4 + 1$ to $p_1 + 2\times p_4 - c$} \label{alg:eq 3 embedding rule}
                \vspace{0.1cm}
				\State // Embedding of the remaining $p_4 - c$ coefficients.
				\vspace{0.1cm}
                     %\begin{align}
                  \Statex \qquad \quad   $y'_{\hati}(k) = y_{\hati}(k - p_4 + c) + \gamma \times t_{\hati}(k - p_1 - p_4 + c)$.  %\nonumber
                     %\end{align}
                     \vspace{0.1cm}
                \EndFor			\label{alg:eq 3 embedding rule end}
                \vspace{0.05cm}
			\EndFor		\\
			\vspace{0.05cm}
			\Return $y'_i$
		\end{algorithmic}
	\end{algorithm*}
}

%%%%%% Stego-image Reconstruction
\subsection{Construction of the Stego-Image}\label{Subsec:L1SABMIS_StegoImageConstruction}
As mentioned earlier, the next step in our scheme is the construction of the stego-image.
Since we can hide a maximum of four secret images into four sub-images of a single cover image, we first construct four sub-stego-images and then perform inverse sampling to obtain a single stego-image. Let $s'_i$ be the sparse vector of the $i^{th}$ block of a sub-stego-image. The sparse vector $s'_i$ is the concatenation of $s'_{i,u}$ and $s'_{i,v}$. Here, the size of ${s}'_{i,u}$, ${s}'_{i,v}$, and ${s}'$ is the same as that of $s_{i,u}$, $s_{i,v}$, and $s$, respectively. Then, we have
%\begin{equation}
\begin{linenomath*}
  \ifcase0% MCS: I find this clearer.
  \begin{align}
    \label{eq:stego reconstruction SABMIS}
    s'_{i,u} &= y'_{i,u}, \refstepcounter{equation} \subeqn \\
    s'_{i,v} &= \argmin_{s'_{i,v}\in \mathbb{R}^{p_2}} \|{s'_{i,v}} \|_{1}
               \quad\text{subject to}\quad
               \Phi s'_{i,v} = y'_{i,v}. \subeqn
               \label{eq:stego reconstruction SABMIS_eq2}
  \end{align}
  \or
\begin{align}\label{eq:stego reconstruction SABMIS}
& s'_{i,u} = y'_{i,u} \: \text{and} \refstepcounter{equation} \subeqn \\
& s'_{i,v} = \argmin_{s'_{i,v}\in \mathbb{R}^{p_2}} \left \|{s'_{i,v}} \right \|_{1} \subeqn \label{eq:stego reconstruction SABMIS_eq2}\\
& \text{Subject to} \hspace{0.1cm} \Phi {s'_{i,v}} = y'_{i,v}, \label{eq:stego reconstruction SABMIS_eq3} \subeqn
\end{align}
  \fi
\end{linenomath*}
%\end{equation}
where $y'_{i}$ is defined in \textbf{Algorithm \ref{alg:Embedding rule SABMIS}}, and it is equal to $\begin{bmatrix}y'_{i,u}\\ y'_{i,v} \end{bmatrix}$ as split in \eqref{eq: propose measurements SABMIS}. 
%$y'_{i,u}$ and $y'_{i,v}$ are defined from \textbf{Algorithm \ref{alg:Embedding rule SABMIS}}, and \eqref{eq: propose measurements SABMIS}. 
The second part \eqref{eq:stego reconstruction SABMIS_eq2} (i.e., obtaining $s'_{i,v}$), is an $\ell_1$-norm minimization problem. Here, we can observe that in the above optimization problem, the constraints are oversampled. As earlier, this oversampling helps us to do sparsification, which leads to increased embedding capacity as well as increased security because  the measurement matrix is encoded. For the solution of the minimization problem \eqref{eq:stego reconstruction SABMIS_eq2}\iffalse--\eqref{eq:stego reconstruction SABMIS_eq3}\fi, we use ADMM \citep{ADMMSBoyd, ADMM_GABAY} to solve the LASSO \citep{Lasso, PeerjLasso} formulation of this minimization problem\footnote{{\color{myyaleblue}Since the linear system of equations in {\eqref{eq:stego reconstruction SABMIS_eq2}\iffalse--\eqref{eq:stego reconstruction SABMIS_eq3}\fi} is overdetermined, we solve it in least squares sense that causes loss of information.}}. We use this method because it has a fast convergence, is easy to implement, and also is extensively used in image processing \citep{ADMMSBoyd, Lasso}.

Next, we convert each vector ${s}'_{i}$ into a block of size $b\times b$. After that, we apply inverse discrete cosine transformation (i.e., the two-dimensional Inverse DCT) to each of these blocks to generate blocks ${x}'_{i}$ of the image. That is,
%After that, we perform inverse sparsification (i.e., we apply the two-dimensional Inverse DCT) to each of these blocks to generate blocks ${x}'_{i}$ of the image. That is,
\begin{linenomath*}
\begin{equation}\label{eq:SABMIS_IDCT}
{x}'_{i}=IDCT\left( {s}'_{i}\right).
\end{equation}
\end{linenomath*}
Next, we construct the sub-stego-image of size $\frac{r}{2} \times \frac{r}{2}$ by arranging all these blocks ${x}'_{i}$. We repeat the above steps to construct all four sub-stego-images. At last, we perform inverse sampling and obtain a single constructed stego-image from these four sub-stego-images. In the experiments section, we show that the quality of the stego-image is also very good. {\color{myyaleblue}The block representation of these steps is given in Figure \ref{Fig:Stego-image construction}. A small numerical example, which further explains this process is given in Appendix \ref{appendix:A small numerical example of working of our stego-image construction process}.}

\begin{figure}[!h]
	\centering
		\includegraphics[width=\textwidth]{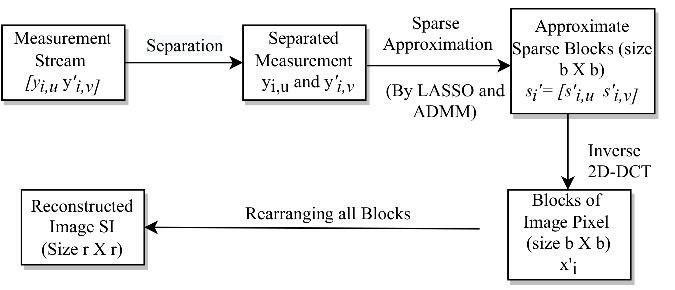}
		\caption{Stego-image construction.}
		\label{Fig:Stego-image construction}
\end{figure}

\subsection{Extraction of the Secret Images}\label{Subsec:L1SABMIS_SecretImagesExtraction}
In this subsection, we discuss the process of extracting secret images from the stego-image. Initially, we perform sampling (as done in Section \ref{Subsec:L1SABMIS_SecretImagesEmbedding} using \eqref{eq:sampling}--\eqref{eq:sampling_eq4}) of the stego-image to obtain four sub-stego-images. Since the extraction of all the secret images is similar, here, we discuss the extraction of only one secret image from one sub-stego-image. {First}, we perform block-wise sparsification of the chosen sub-stego-image. For this, we divide the sub-stego-image into blocks of size $b\times b$. We obtain a total of $\frac{r^2}{4\times b^2}$ blocks. Further, we sparsify each block (say $x''_i$) by computing the corresponding sparse vector (say~$s''_i$). That is,
\begin{linenomath*}
\begin{equation}
%s''_i = \Psi^T x_i,
s''_i = DCT(x''_i).
\end{equation}
\end{linenomath*}

Next, as earlier, we arrange these sparse blocks in a zig-zag scanning order, obtain the corresponding sparse vectors each of size $b^2 \times 1$, and then categorize each of them into two groups $s''_{i,u}\in \mathbb{R}^{p_1}$ and $s''_{i,v}\in \mathbb{R}^{p_2}$. Here, as before, $p_1$ and $p_2$ are the numbers of coefficients having large values and small values (or zero values), respectively. After that, we oversample each sparse vector using linear measurements (say $y''_i\in \mathbb{R}^{(p_1+p_3)\times 1}$),
%as done in subsection \ref{Subsec:L1SABMIS_SecretImagesEmbedding}, using the same measurement matrix $\Phi\in \mathbb{R}^{m\times p_2}$ and \eqref{eq: propose measurements SABMIS}.
\begin{linenomath*}
\begin{equation}
y''_i=\begin{bmatrix}y''_{i,u}\\ y''_{i,v} \end{bmatrix}=
\begin{bmatrix}
s''_{i,u}\\ \Phi s''_{i,v}
\end{bmatrix}.
\end{equation}
\end{linenomath*}
From $y''_i$, we extract the DCT coefficients of the embedded secret image using \textbf{Algorithm \ref{alg:Extraction rule SABMIS}}. This extraction rule is the reverse of the embedding rule given in \textbf{Algorithm \ref{alg:Embedding rule SABMIS}}.
%...................Embeextraction Rule Algorithm.......................%
{
	\begin{algorithm*}[!t]
	%\scriptsize %\small, \footnotesize, \scriptsize, or \tiny
	%\footnotesize
		\caption{Extraction Rule}\label{alg:Extraction rule SABMIS}
		\begin{algorithmic}[1]
			\renewcommand{\algorithmicrequire}{\textbf{Input:}}
			\renewcommand{\algorithmicensure}{\textbf{Output:}}
			\Require \quad
			\begin{itemize}
				\item ${y''_i}$: Sequence of linear measurements of the stego-image with $i=1,2,\hdots,\frac{r^2}{4\times b^2}$.
				%\item $p_1$ and $p_4$ are lengths of certain vectors defined on pages \pageref{p1_page_ref_section} and \pageref{p4_page_ref_section}, respectively.
				%\item $\alpha$, $\beta$, $\gamma$, and $c$ are algorithmic constants that are chosen based upon experience. The choices of these constants are discussed in the experimental results section.
				\item {\color{myyaleblue}$p_1$, $p_4$, $\alpha$, $\beta$, $\gamma$, and $c$ are chosen as in \textbf{Algorithm \ref{alg:Embedding rule SABMIS}}.}
			\end{itemize}

			\Ensure \quad
			\begin{itemize}
				%\item $t'$: Sequence of transform coefficients of the extracted secret image.
				\item $t'_{\hati}$: Sequence of transform coefficients of the secret image with $\hati=1,2,\hdots,\frac{m^2}{l^2}$.
			\end{itemize}
		%	\\ \textit{Initialisation}:
		    \State Initialize $t'_{\hati}$ to zeros, where $\hati=1,2,\hdots,\frac{m^2}{l^2}$. % and $\frac{m^2}{l^2}$ will be less than or equal to $\frac{r^2}{4\times b^2}$ (discussed on page \pageref{page_coeffm2/l2}.
			\For {${\hati} = 1$ to $\frac{m^2}{l^2}$} 
			\vspace{0.1cm}
				\State // Extraction of the first coefficient.
			\vspace{0.15cm}
              \Statex    \qquad \hspace{-0.3cm} ${t'}_{\hati}(1)  = \frac{y''_{\hati}(p_1) - y''_{\hati}(p_1-2c)}{\alpha}$.  \nonumber
            \vspace{0.1cm}
                \For {$j = p_1 - c + 1$ to $p_1 - 1$}
            \vspace{0.1cm}
 	                \State // Extraction of the next $c-1$ coefficients.
			\vspace{0.15cm}
                 \Statex  \qquad \quad ${t'_{\hati}}(j-p_1+c+1)  = \frac{y''_{\hati}(j) - y''_{\hati}(j-c)}{\beta}$.  \nonumber
			\vspace{0.1cm}
                \EndFor
                \For {$k = p_1 + p_4 + 1$ to $p_1 + 2\times p_4 - c$}
             \vspace{0.1cm}
	                \State // Extraction of the remaining $p_4 - c$ coefficients.
             \vspace{0.15cm}
                   \Statex  \qquad \quad ${t'_{\hati}}(k - p_1 - p_4 + c)  = \frac{y''_{\hati}(k) - y''_{\hati}(k - p_4 + c)}{\gamma}$.  \nonumber
            \vspace{0.1cm}
                \EndFor
            \vspace{0.05cm}
			\EndFor		\\
		\vspace{0.05cm}
			\Return ${t'_{\hati}}$
		\end{algorithmic}
	\end{algorithm*}
}

In \textbf{Algorithm \ref{alg:Extraction rule SABMIS}}, $t'_{\hati} \in \mathbb{R}^{l^2\times 1}$, for ${\hati}=1,2,\hdots,\frac{m^2}{l^2}$, are the vector representations of the DCT coefficients of the blocks of one extracted secret image. Next, we convert each vector $t'_{\hati}$ into blocks of size $l\times l$, and then perform a block-wise Inverse DCT (IDCT) (using \eqref{eq:SABMIS_IDCT}) to obtain the secret image pixels. Finally, we obtain the extracted secret image of size $m\times m$ by arranging all these blocks column wise. As mentioned earlier, this steganography scheme is a blind multi-image steganography scheme because it does not require any cover image data at the receiver side for the extraction of secret images.

{\color{myblue}Here, the process of hiding (and extracting) secret images is not fully lossless\footnote{This is common in transform-based image steganography.}, resulting in the degradation of the quality of extracted secret images. This is because we first oversample the original image using \eqref{eq: propose measurements SABMIS}, and then we construct the stego-image by solving the optimization problem \eqref{eq:stego reconstruction SABMIS_eq2}, which leads to a loss of information.} However, our algorithm is designed in such a way that we are able to extract high-quality secret images. We support this fact with examples in the experiments section (specifically, Section \ref{secretImageQuality}). We term our algorithm Sparse Approximation Blind Multi-Image Steganography (SABMIS) scheme due to the involved sparse approximation and the blind multi-image steganography.

\begin{figure}[!h]
	\centering
		\includegraphics[width=\textwidth]{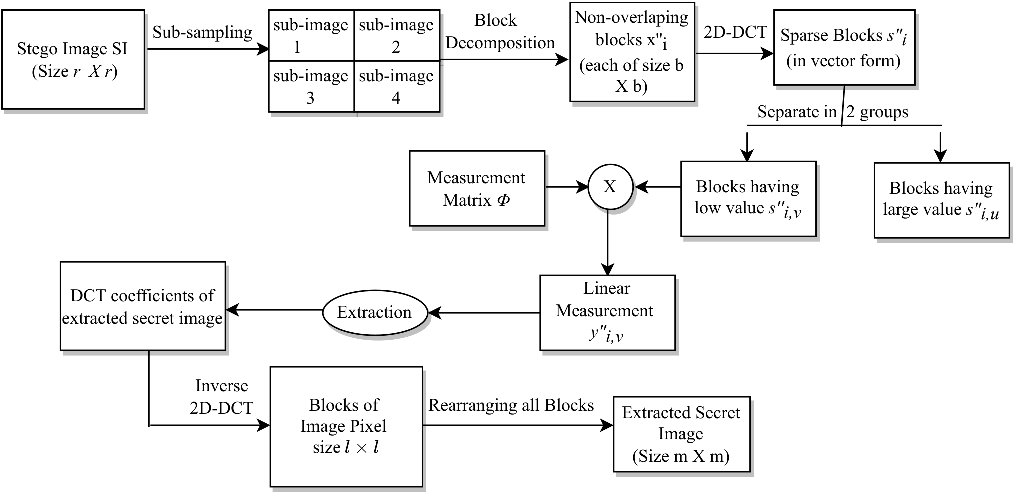}
		\caption{The extraction process.}
		\label{Fig:The extraction process}
\end{figure}

{\color{myyaleblue}The above extraction process is represented via a block diagram in Figure \ref{Fig:The extraction process}. As discussed earlier, this extraction is just the reverse of the embedding process.%, and due to space constraints, we do not add a numerical example for this.
%A small numerical example, which further enlighten this process is given in Appendix \ref{appendix:A small numerical example of working of our secret image extraction process}.
}

\section{Experimental Results} \label{Sec:L1SABMIS_NumericalExp}
Experiments are carried out in MATLAB on a machine with an Intel Core i5 processor @2.50 GHz and 8GB RAM. We use 10 standard test images (those which are frequently found in literature) for our experiments. These image are freely available with no copyright \citep{standard-test-images-for-Image-Processing}.

\begin{figure}[!h]
	\centering
		\includegraphics[width=\textwidth]{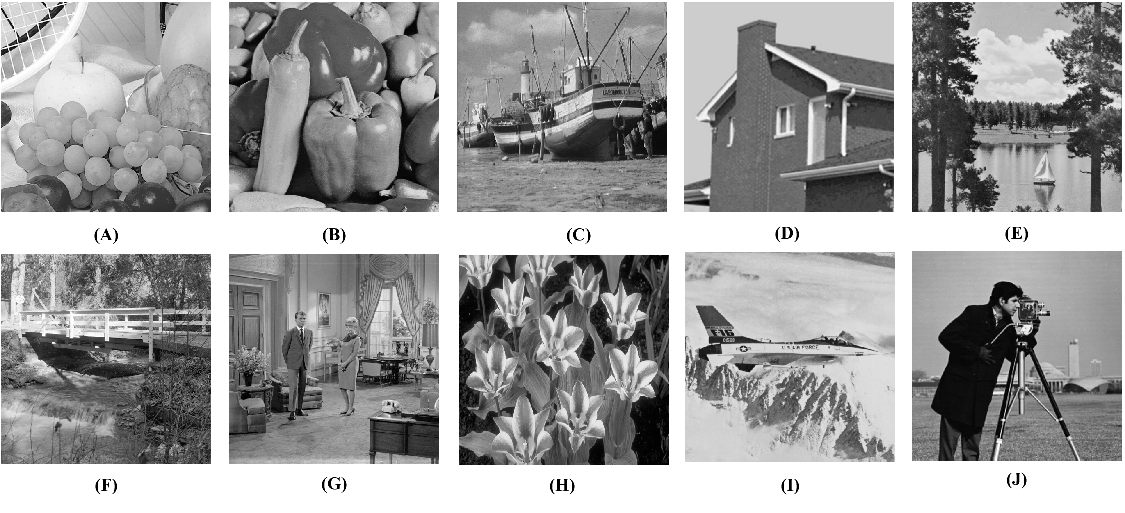}
		\caption{Test images used in our experiments \citep{standard-test-images-for-Image-Processing}. (A) Fruits, (B) Peppers, (C) Boat, (D) House, (E) Lake, (F) Stream, (G) Living room, (H) Tulips, (I) Airplane, and (J) Camera man. \url{https://github.com/mohammadimtiazz/standard-test-images-for-Image-Processing}}
		\label{Figure:test images SABMIS}
\end{figure}

Here, we take all ten images shown in Figure \ref{Figure:test images SABMIS} as the cover images, and four images; Figures 6(B), 6(E), 6(F), and 6(J) as the secret images for our experiments. However, we can use any of the ten images as the secret images.

% Cover image
Although the images shown in Figure \ref{Figure:test images SABMIS} look to be of the same dimension, they are of varying sizes. For our experiments, each cover image is converted to $1024\times 1024$ size (i.e., $r\times r$). We take blocks of size $8\times 8$ for the cover images (i.e., $b\times b$). Recall from subsection \ref{Subsec:L1SABMIS_SecretImagesEmbedding} that the size of the DCT sparsified vectors is $(p_1+p_2) \times 1$ with $p_1+p_2=b^2$ (here, $b^2 = 64$). In general, applying DCT on images results in sparse vectors where more than half of the coefficients have values that are either very small or zero \citep{Rohit_CSIS, Steg_Rohit, Pan}. This is the case here as well. Hence, in our experiments, we take $p_1 = p_2 = 32$. Recall, the size of the measurement matrix $\Phi$ is $p_3\times p_2$ with $p_3 > p_2$. We take $p_3 = 50\times p_2$. Without loss of generality, the element values of the column-normalized measurement matrix are taken as random numbers with mean $0$ and standard deviation $1$, which is a common standard.

% Secret Image
{\color{myrosewood}There are many options for taking the size of the secret images. In one way the size of the length and the width of the secret image is taken to be the same as the length and the width of the cover image \citep{Sanjutha_MK}. Another approach, which many papers follow, the dimensions of secret image is taken to be substantially smaller than the dimensions of the cover image. For example, the size of the length and the width of the secret image to be half of the length and the width of the cover image \citep{S_Hemalatha, S_Arunkumar_MedicalImage_JIFS, S_Arunkumar_MedicalImage_DWT_2019}, respectively. Another option is to use a factor of one-fourth \citep{Ste_2014_4hide}. Hence, without any loss of generality, we take the dimensions of secret image to be half of the dimensions of cover image.}

Thus, each of the secret image is converted to $512\times 512$ size (i.e., $m\times m$). 
%This choice is also motivated by the fact that we chose the size of the secret image to be half of that of the cover image ($1024\times 1024$). 
We take blocks of size $8\times 8$ for the secret images as well (i.e., $l\times l$). In general, the DCT coefficients can be divided into three sets \citep{shastri2018density}; low frequencies, middle frequencies, and high frequencies. Low frequencies are associated with the illumination, middle frequencies are associated with the structure, and high frequencies are associated with the noise or small variation details. Thus, these high-frequency coefficients are of very little importance for the to-be embedded secret images. Since the number of high-frequency coefficients is usually half of the total number of coefficients, we take $p_4 = 32$ (using $8\times 8$ divided by $2$).

%We embed these coefficients of the secret image using \textbf{Algorithm \ref{alg:Embedding rule SABMIS}}
{\color{myblue}The values of the constants in \textbf{Algorithm \ref{alg:Embedding rule SABMIS}} and \textbf{Algorithm \ref{alg:Extraction rule SABMIS}} are taken as follows\footnote{The values of these constants do not affect the convergence of ADMM much. Determining the range of values that work best here is part of our future work.} (based upon experience): $\alpha=0.01$, $\beta=0.1$, $\gamma=1$, and $c=6$. 
The LASSO constant is taken as $\lambda = 0.011 \lambda_{max}$, where $\lambda_{max} = \|\Phi^T y'_{i,v}\|_{\infty}$ with $\Vert \cdot \Vert_{\infty}$ being the $\ell_{\infty}$-norm \citep{Rohit2021apache}. For ADMM, we set %the maximum number of iterations as $500$, 
the absolute stopping tolerance as $1\times 10^{-4}$, and the relative stopping tolerance as $1\times 10^{-2}$. These values are again taken based upon our experience with a similar algorithm \citep{Rohit2021apache}. Eventually, our ADMM always converges in $5$ to $20$ iterations.

As mentioned earlier, in the {\color{myrosewood}six} sections below we experimentally demonstrate the usefulness of our {steganography scheme}. In Section \ref{Embedding Capacity}, we show analytically that our SABMIS scheme gives excellent embedding capacities. In Section \ref{Stego-image Quality Assessment st2}, we show that the quality of the constructed stego-images, when compared with the corresponding cover images, is high. In Section \ref{secretImageQuality}, we demonstrate the good quality of the extracted secret images when compared with the original secret images. In Section \ref{Security Analysis st2}, we show that our SABMIS scheme is resistant to steganographic attacks. 
{\color{myrosewood}In Section \ref{Taming Data}, we demonstrate efficiency of SABMIS by providing its timing data.} 
%In Section \ref{Performance Comparison SABMIS}, we perform a comparison of our scheme with competing steganography schemes as discussed in the introduction.}
{\color{myyaleblue}{In Section \ref{Application of our scheme on real-life data}, we discuss applicability of our scheme to real-life data, and hence, demonstrate its practical usefulness.}}

\subsection{Embedding Capacity Analysis}\label{Embedding Capacity}
The embedding capacity (or embedding rate) is the number (or length) of secret bits that can be hidden/ embedded in each pixel of the cover image. It is measured in bits per pixel\footnote{Since in the transform domain-based steganography schemes, some specific transform coefficients are hidden into the cover image (along with the secret bits), a more appropriate term that can be used for embedding capacity is ``bits of information per pixel'' (bipp). However, to avoid confusion, we use the term \si{bpp} in this paper, which is commonly used.} (bpp) and is calculated as follows:
\begin{linenomath*}
\begin{equation}
\text{EC in \si{bpp}} = \frac{\text{Total number of secret bits embedded}}{\text{Total number of pixels in the cover image}}.
\end{equation}
\end{linenomath*}
{\color{myrosewood}As motivated on the previous page, we chose the size of the length and the width of secret image to be half of the length and the width of cover image, respectively. Since our cover images are of size $1024\times 1024$, our secret images are taken to be of size $512\times 512$. For a grayscale image, each pixel size is \SI{8} bits. Hence, when hiding one secret image in a cover image, we obtain embedding capacity as below.}
\begin{linenomath*}
\begin{equation}
\text{EC in \si{bpp}} = \frac{{512 \times 512 \times 8}}{{1024 \times 1024}}, 
\end{equation}
\end{linenomath*}
{\color{myrosewood}which is equal to \SI{2} bpp. Similarly, while hiding two, three, and four secret images in a cover image, we obtain the embedding capacities of \SI{4} bpp, \SI{6} bpp, and \SI{8} bpp, respectively.}

\subsection{Stego-Image Quality Assessment}\label{Stego-image Quality Assessment st2}
In general, the visual quality of the stego-image degrades as the embedding capacity increases. Hence, preserving the visual quality becomes increasingly important. There is no universal criterion to determine the quality of the constructed stego-image. However, we evaluate it by visual and numerical measures.
We use Peak Signal-to-Noise Ratio (PSNR), Mean Structural Similarity (MSSIM) index, Normalized Cross-Correlation (NCC) coefficient, entropy, and Normalized Absolute Error (NAE) numerical measures.

%\subsubsection{Subjective or Visual Measure}\label{Visual similarity}
When using the visual measures, we construct the stego-images corresponding to the different {\color{myblue}cover} images used in our experiments and then check their distortion visually. We also check their corresponding edge map diagrams. Here, we present the visual comparison only for `Stream' as the cover image with {\color{myblue}`Lake' secret image and the corresponding stego-image}. We get similar results for the other images as well. The comparison is given in Figure \ref{Figure:visual analysis st2}. The cover image and its corresponding edge map are shown in parts (a) and (b) of this figure. The stego-image and its corresponding edge map are given in parts (c) and (d) of the same figure. When we compare each figure with its counterpart, we find that they are very similar.

\begin{figure}[!h]
	\centering
		\includegraphics[width=\textwidth]{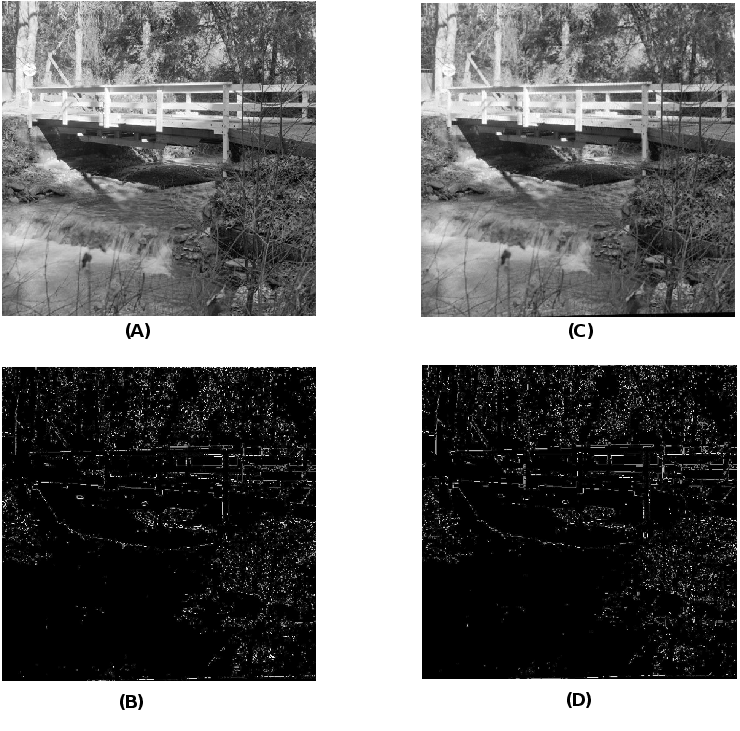}
		\caption{Visual quality analysis between `Stream' cover image (CI) and its corresponding stego-image (SI) (with `Lake' secret image embedded in it). (A) cover image, (B) cover image edge map, (C) stego-image, and (D) stego-image edge map.}
	\label{Figure:visual analysis st2}
\end{figure}

%\begin{figure}[!h]
%	\centering
%	\begin{multicols}{2}
%	\begin{subfigure}[b]{0.40\textwidth}
%		\centering
%		\includegraphics[width=5.5cm,height=5.5cm,keepaspectratio]{Stream.eps}
%		\caption{Cover Image (CI)}
%		\label{fig:stream cover image st2}
%	\end{subfigure}
%	%\hfill %
%	\begin{subfigure}[b]{0.40\textwidth}
%		\centering
%		\includegraphics[width=5.5cm,height=5.5cm,keepaspectratio]{StreamCoverEdgeMap_st2.eps}
%		\caption{CI Edge Map}
%		\label{fig:stream cover edge map st2}
%	\end{subfigure}
%	%\hfill %
%	\begin{subfigure}[b]{0.40\textwidth}
%		\centering
%		\includegraphics[width=5.5cm,height=5.5cm,keepaspectratio]{StegoImage_st2_stream.eps}
%		\caption{Stego-Image (SI)}
%		\label{fig:stream stego-image st2}
%	\end{subfigure}
%%\hfill %
%	\begin{subfigure}[b]{0.40\textwidth}
%		\centering
%		\includegraphics[width=5.5cm,height=5.5cm,keepaspectratio]{StreamStegoEdgeMap_st2.eps}
%		\caption{SI Edge Map}
%		\label{fig:stream stego edge map st2}
%	\end{subfigure}
%	%\caption{Edge map of `Pepper' cover image and its stego-image for parameter $|p_1|$=12 and $|m|$=37.}
%	%\label{Fig:Edge map}
%	%\hfill %
%	\end{multicols}
%	\caption{Visual quality analysis between `Stream' cover image (CI) and its corresponding stego-image (SI) (with `Lake' secret image embedded in it).}
%	\label{Figure:visual analysis st2}
%\end{figure}

{\color{myblue}Next, when using the numerical measures to assess the quality of the stego-image with respect to the cover image, we first evaluate the most common measure of PSNR value in Section \ref{PSNRSec}. Subsequently, we evaluate the other more rarely used numerical measures of MSSIM index, NCC coefficient, entropy, and NAE in Section \ref{OtherMeasuresSec}}.

\subsubsection{Peak Signal-to-Noise Ratio (PSNR) Value}\label{PSNRSec}
%\paragraph{\textbf{PSNR:}}\label{PSNR}
We compute the \textit{PSNR} values to evaluate the imperceptibility of stego-images (SI) with respect to the corresponding cover images (CI) as follows \citep{PeerjPSNRSSIM}:
\begin{linenomath*}
\begin{equation}%\label{eq:PSNR}
PSNR(CI, SI)=10\log_{10}\frac{R^{2}}{MSE(CI, SI)}\: \si{dB},
\end{equation}
\end{linenomath*}
where $R$ is the maximum intensity of the pixels, which is $255$ for grayscale images, \si{dB} refers to decibel, and $MSE(CI, SI)$ represents the mean square error between the cover image $CI$ and the stego-image $SI$ that is calculated as
\begin{linenomath*}
\begin{equation}%\label{eq:MSE}
MSE(CI, SI)=\frac{\sum_{i=1}^{r1}\sum_{j=1}^{r2}\left ( CI\left ( i,\, j \right )-SI\left ( i,\, j \right ) \right )^2}{r1\times r2},
\end{equation}
\end{linenomath*}
where $r1$ and $r2$ represent the row and column numbers of the image (for us either cover or stego), respectively, and $CI(i,j)$ and  $SI(i,j)$ represent the pixel values of the cover image and the stego-image, respectively.

A higher PSNR value indicates a higher imperceptibility of the stego-image with respect to the corresponding cover image. In general, a value higher than \SI{30}{dB} is considered to be good since human eyes can hardly distinguish the distortion in the image \citep{Ste_2020, Zhang2013, Liu}.

The PSNR values of the stego-images corresponding to the ten cover images are given in Figure \ref{fig:PSNRStego1imagehidden} and Figure \ref{fig:PSNRStego1to4Imageshidden}. In Figure \ref{fig:PSNRStego1imagehidden}, we show the PSNR values of all the stego-images when separately all the four secret images (mentioned above in Figure \ref{Figure:test images SABMIS}) are hidden. In this figure, we obtain the highest PSNR value (\SI{46.25}{dB}) when the `Peppers' secret image is hidden in the `House' cover image, while the lowest PSNR value (\SI{37.66}{dB}) is obtained when the `Stream' secret image is hidden in the `Stream' cover image.

In Figure \ref{fig:PSNRStego1to4Imageshidden}, we show the PSNR values for the four cases of hiding one, two, three, and four secret images in the ten cover images. As we have four secret images, when hiding one secret image, we have a choice of hiding any one of them and present the resulting PSNR values. However, we separately hide all four images, obtain their PSNR values, and then present the average results. Similarly, the average PSNR values are presented for the cases when we hide two and three images. In this figure, we obtain the highest average PSNR value (\SI{45.21}{dB}) when one secret image is hidden in the `House' cover image, while the lowest PSNR value (\SI{31.78}{dB}) is obtained when all four secret images are hidden in the `Stream' cover image. Also, we observe that for all test cases, we obtain PSNR values higher than \SI{30}{dB} which, as earlier, are considered good.

\par
\begin{figure}[h]
	\centering
			\includegraphics[width=1\textwidth]{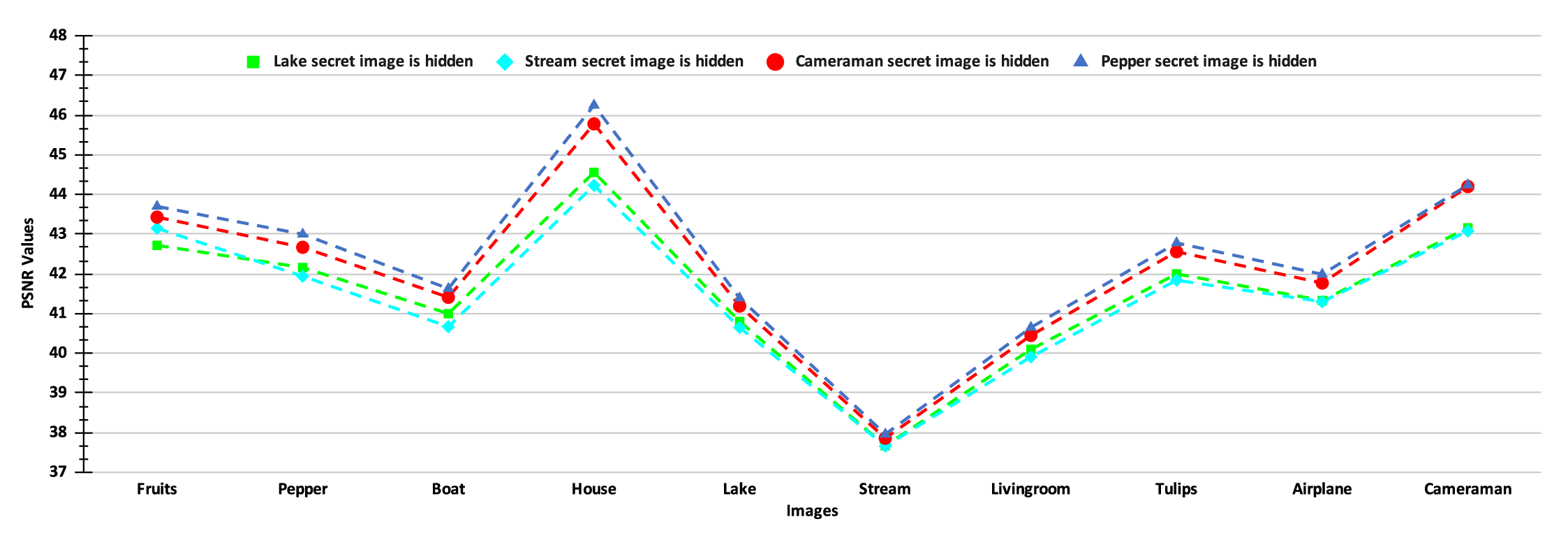}
		\caption{PSNR values of the stego-images when only one secret image is hidden.}
		\label{fig:PSNRStego1imagehidden}
\end{figure}

\begin{figure}[!h]
	\centering
		\includegraphics[width=1\textwidth]{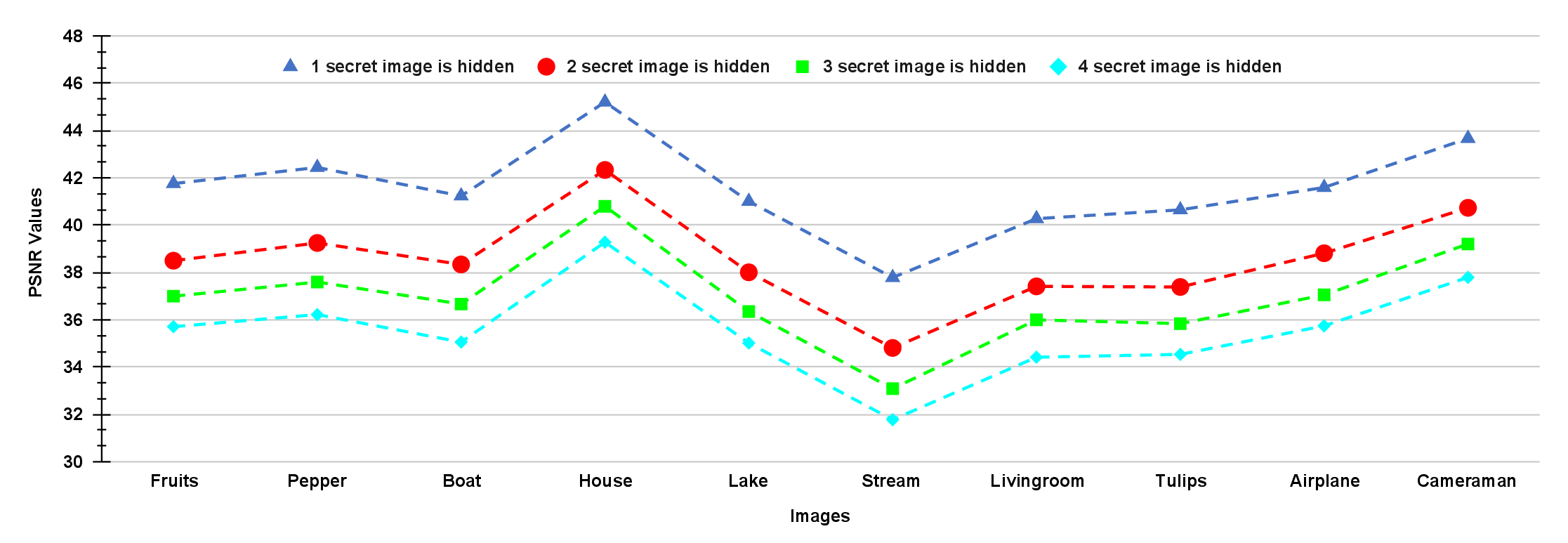}
		\caption{PSNR values of the stego-images when different numbers of images are hidden.}
		\label{fig:PSNRStego1to4Imageshidden}
\end{figure}

\subsubsection{Other Numerical Measures}\label{OtherMeasuresSec}
\quad
\paragraph{\textbf{Mean Structural Similarity (MSSIM) Index}}\label{SSIM}
This is an image quality assessment metric used to measure the structural similarity between {\color{myblue}two images, which is most noticeable to humans \citep{SSIM, PeerjPSNRSSIM}. MSSIM between the cover image $CI$ and the stego-image $SI$ is given as
%%%%%%%%%%%%%%%%%%% SSIM equation %%%%%%%%%%%%%%%%%
\begin{linenomath*}
\begin{align}%\label{eq:SSIM}
& MSSIM(CI, SI)=\frac{1}{M}\sum_{j=1}^{M}SSIM(ci_{j}, si_{j}),
\end{align} %%%%%%%%%%%%%%%%%%%%%%%%%%%%%%%%%%%%%%%
\end{linenomath*}
where $ci_{j}$ and $si_{j}$ are the pixel values of the cover image and the stego-image, respectively, at the $j^{th}$ local window\footnote{It is a 11×11 Gaussian matrix, which is standard in the calculation of MSSIM.} with $M$ being the number of local windows \citep{SSIM, SSIM2}, and
\begin{linenomath*}
\begin{align}%\label{eq:SSIM}
& SSIM(x, y)=\frac{(2\mu _{x}\mu _{y}+C_{1})(2\sigma _{xy}+C_{2})}{(\mu _{x}^{2}+\mu _{y}^{2}+C_{1})(\sigma _{x}^{2}+\sigma _{y}^{2}+C_{2})},\end{align} %%%%%%%%%%%%%%%%%%%%%%%%%%%%%%%%%%%%%%%
\end{linenomath*}
where for vectors $x$ and $y$; $\mu_{x}$ is the weighted mean of $x$; $\mu_{y}$ is the weighted mean of $y$; $\sigma_{x}$ is the weighted standard deviation of $x$; $\sigma_{y}$ is the weighted standard deviation of $y$; $\sigma_{xy}$ is the weighted covariance between $x$ and $y$; $C_{1}$ and $C_{2}$ are positive constants.

We take $M$ = $1069156$, $C_1 = (0.01\times255)^2$, and $C_2 = (0.03\times255)^2$ based upon the recommendations from \citep{SSIM, SSIM2}.
The value of the MSSIM index lies between $0$ and~$1$, where the value $0$ indicates that there is no structural similarity between the cover image and the corresponding stego-image, and the value $1$ indicates that the images are identical.}
%\textcolor{red}{Correct?}

\paragraph{\textbf{Normalized Cross-Correlation (NCC) Coefficient:}}\label{NNC}
{\color{myblue}This metric measures the amount of correlation between two images \citep{NC}. The NCC coefficient between the cover image $CI$ and the stego-image $SI$ is given as
%%%%%%%%%%%%%%%%%%%%% Normalized cross-correlation equation
\begin{linenomath*}
\begin{align}%\label{eq:NC}
& NCC(CI,SI)=\frac{\sum_{i=1}^{r1}\sum_{j=1}^{r2}CI(i,j)SI(i,j)}{\sum_{i=1}^{r1}\sum_{j=1}^{r2}CI^{2}(i,j)},
\end{align} %%%%%%%%%%%%%%%%%%%%%%%%%%%%%%%%%%%%%%%
\end{linenomath*}
where $r1$ and $r2$ represent the row and column numbers of the image (for us either cover or stego), respectively, and $CI(i,j)$ and $SI(i,j)$ represent the pixel values of the cover image and the stego-image, respectively. The NCC coefficient value of $0$ indicates that the cover image and the stego-image are not correlated while a value of $1$ indicates that the two are highly correlated.}
%\textcolor{red}{Correct?}

\paragraph{\textbf{Entropy:}}\label{Entropy}
In general, entropy is defined as the measure of average uncertainty of a random variable.
% , which here is the average number of bits required to describe the random variable.
In the context of an image, it is a statistical measure of randomness that can be used to characterize the texture of the image \citep{Gonzalez}. For a grayscale image {\color{myblue}(either a cover image or a stego-image in our case)}, entropy is given as
%%%%%%%%%%%%%%%%%%%%% Entropy equation %%%%%%%%%%%%%%
\begin{linenomath*}
\begin{align}%\label{eq:entropy}
  E=-\sum_{i=0}^{255}(p_{i}\log _{2}p_{i}), % MCS: better?
  %Entropy=-\sum_{i=0}^{255}(p_{i}\log _{2}p_{i}),
\end{align} %%%%%%%%%%%%%%%%%%%%%%%%%%%%%%%%%%%%%%%
\end{linenomath*}
where $p_{i} \in [0, 1]$ is the fraction of image pixels that have the value~$i$. If the stego-image is similar to its corresponding cover image, then the two should have similar entropy values (due to similar textures).

\paragraph{\textbf{Normalized Absolute Error (NAE):}}\label{NAE}
This metric is a distance measure that captures pixel-wise differences between two images \citep{S_Arunkumar_MedicalImage_DWT_2019}. NAE between the cover image $CI$ and the stego-image $SI$ is given as
\begin{linenomath*}
\begin{equation}%\label{eq:NAE}
NAE(CI, SI)=\frac{\sum_{i=1}^{r1}\sum_{j=1}^{r2}\left ( | CI\left ( i,\, j \right )-SI\left ( i,\, j \right ) | \right )}{\sum_{i=1}^{r1}\sum_{j=1}^{r2} CI\left ( i,\, j \right )},
\end{equation}
\end{linenomath*}
where $r1$ and $r2$ represent the row and the column numbers of the image (for us either cover or stego), respectively, and $CI(i,j)$ and $SI(i,j)$ represent the pixel values of the cover image and the stego-image, respectively. NAE has values in the range $0$ to $1$. A value close to $0$ indicates that the cover image is very close to its corresponding stego-image, and a value close to $1$ indicates that the two are substantially far apart.

\par
In Table \ref{Table:Stego-Images Quality SABMIS}, we present the values of MSSIM index, NCC coefficient, entropy and NAE for our SABMIS scheme when hiding all four secret images. We do not present the values for the cases of embedding less than four secret images as their results will be better than those given in Table \ref{Table:Stego-Images Quality SABMIS}. Hence, our reported results are for the worst case. From this table, we observe that all values of the MSSIM index are {\color{myblue}nearly} equal to $1$ {\color{myblue}(different in the sixth place of decimal)}, the values of NCC coefficients are close to 1, and values of NAE are close to $0$. The entropy values of the cover and the stego-images are almost identical. All these values indicate that the cover images and their corresponding stego-images are almost identical.
\par
%%%%%%%%%%%%%%%%%%Stego-Images Quality Table%%%%%%%%%%%%%%%%%%
\begin{table}[!h]
\centering
\footnotesize
\caption{MSSIM index, NCC coefficient, entropy, and NAE of the stego-images when compared with the corresponding cover images.}\label{Table:Stego-Images Quality SABMIS}
\setlength{\tabcolsep}{10pt}
\begin{tabular}{cccccc}
\hline
\multicolumn{1}{|c|}{\multirow{2}{*}{\textbf{\begin{tabular}[c]{@{}c@{}}Cover\\ Image\end{tabular}}}} & \multicolumn{1}{c|}{\multirow{2}{*}{\textbf{MSSIM}}} & \multicolumn{1}{c|}{\multirow{2}{*}{\textbf{NCC}}}  & \multicolumn{2}{c|}{\textbf{Entropy}}                                                                                                                & \multicolumn{1}{c|}{\multirow{2}{*}{\textbf{NAE}}} \\ \cline{4-5}

\multicolumn{1}{|c|}{}  & \multicolumn{1}{c|}{}  & \multicolumn{1}{c|}{}                               & \multicolumn{1}{c|}{\textbf{\begin{tabular}[c]{@{}c@{}}Cover\\ Image\end{tabular}}} & \multicolumn{1}{c|}{\textbf{\begin{tabular}[c]{@{}c@{}}Stego-\\ Image\end{tabular}}} & \multicolumn{1}{c|}{} \\ \hline

\multicolumn{1}{|c|}{Fruits}                                                                            & \multicolumn{1}{c|}{1}                               & \multicolumn{1}{c|}{0.9996}                                                & \multicolumn{1}{c|}{7.488}                         & \multicolumn{1}{c|}{7.496} & \multicolumn{1}{c|}{0.009}                                                            \\ \hline
\multicolumn{1}{|c|}{Peppers}                                                                          & \multicolumn{1}{c|}{1}                               & \multicolumn{1}{c|}{0.9997}                                                & \multicolumn{1}{c|}{7.573}                                                             & \multicolumn{1}{c|}{7.603} & \multicolumn{1}{c|}{0.012}                                                           \\ \hline
\multicolumn{1}{|c|}{Boat}                                                                            & \multicolumn{1}{c|}{1}                               & \multicolumn{1}{c|}{0.9998}                                                & \multicolumn{1}{c|}{7.121}                                                             & \multicolumn{1}{c|}{7.151}      & \multicolumn{1}{c|}{0.012}                                                      \\ \hline
\multicolumn{1}{|c|}{House}                                                                        & \multicolumn{1}{c|}{1}                               & \multicolumn{1}{c|}{0.9998}                                                 & \multicolumn{1}{c|}{5.756}                                                              & \multicolumn{1}{c|}{6.630}   & \multicolumn{1}{c|}{0.007}                                                        \\ \hline
\multicolumn{1}{|c|}{Lake}                                                                           & \multicolumn{1}{c|}{1}                               & \multicolumn{1}{c|}{0.9997}                                                & \multicolumn{1}{c|}{7.471}                                                             & \multicolumn{1}{c|}{7.513}                  & \multicolumn{1}{c|}{0.013}                                           \\ \hline
\multicolumn{1}{|c|}{Stream}                                                                         & \multicolumn{1}{c|}{1}                               & \multicolumn{1}{c|}{0.9991}                                                & \multicolumn{1}{c|}{7.702}                                                             & \multicolumn{1}{c|}{7.719}             & \multicolumn{1}{c|}{0.020}                                               \\ \hline
\multicolumn{1}{|c|}{Livingroom}                                                                      & \multicolumn{1}{c|}{1}                               & \multicolumn{1}{c|}{0.9996}                                                & \multicolumn{1}{c|}{7.431}                                                             & \multicolumn{1}{c|}{7.438}                & \multicolumn{1}{c|}{0.014}                                            \\ \hline
\multicolumn{1}{|c|}{Tulips}                                                                            & \multicolumn{1}{c|}{1}                               & \multicolumn{1}{c|}{0.9994}                                                & \multicolumn{1}{c|}{7.713}                                                             & \multicolumn{1}{c|}{7.735}              & \multicolumn{1}{c|}{0.011}                                              \\ \hline
\multicolumn{1}{|c|}{Jetplane}                                                                        & \multicolumn{1}{c|}{1}                               & \multicolumn{1}{c|}{0.9998}                                                & \multicolumn{1}{c|}{6.716}                                                              & \multicolumn{1}{c|}{6.795}     & \multicolumn{1}{c|}{0.008}                                                       \\ \hline
\multicolumn{1}{|c|}{Cameraman}                                                                       & \multicolumn{1}{c|}{1}                               & \multicolumn{1}{c|}{0.9999}                                                    & \multicolumn{1}{c|}{7.055}                                                             & \multicolumn{1}{c|}{7.133}                  & \multicolumn{1}{c|}{0.009}                                           \\ \hline
\multicolumn{1}{|c|}{\textbf{Average}}                                                                & \multicolumn{1}{c|}{\textbf{1}}                      & \multicolumn{1}{c|}{\textbf{0.9996}}                             & \multicolumn{1}{c|}{\textbf{7.202}}   & \multicolumn{1}{c|}{\textbf{7.320}}                                                 & \multicolumn{1}{c|}{\textbf{0.011}} \\ \hline
\end{tabular}
\end{table}
\par

%%%%%%%% Secret Image Quality %%%%%%%%%%%%%
\subsection{Secret Image Quality Assessment}\label{secretImageQuality}
%{\color{myblue}We have demonstrated that our stego-image is of good quality. Next, we demonstrate that the extracted secret images are also very good.}
Since human observers are considered the final arbiter to assess the quality of the extracted secret images, we compare one such original secret image and its corresponding extracted secret image. The results of all other combinations are almost the same. In Figures 10(A) and 10(C), we show the original `Lake' secret image and the extracted `Lake' secret image (from the `Stream' stego-image). From these figures, we observe that there is little distortion in the extracted image. Besides this, for these two images, we also present their corresponding edge map diagrams (in Figures 10(B) and 10(D), respectively). Again, we observe minimal variations between the original and the extracted secret images.
\par

%%%%%%%%%%%%%%% Figures %%%%%%%%%%%%%
\begin{figure}[!h]
	\centering
		\includegraphics[width=\textwidth]{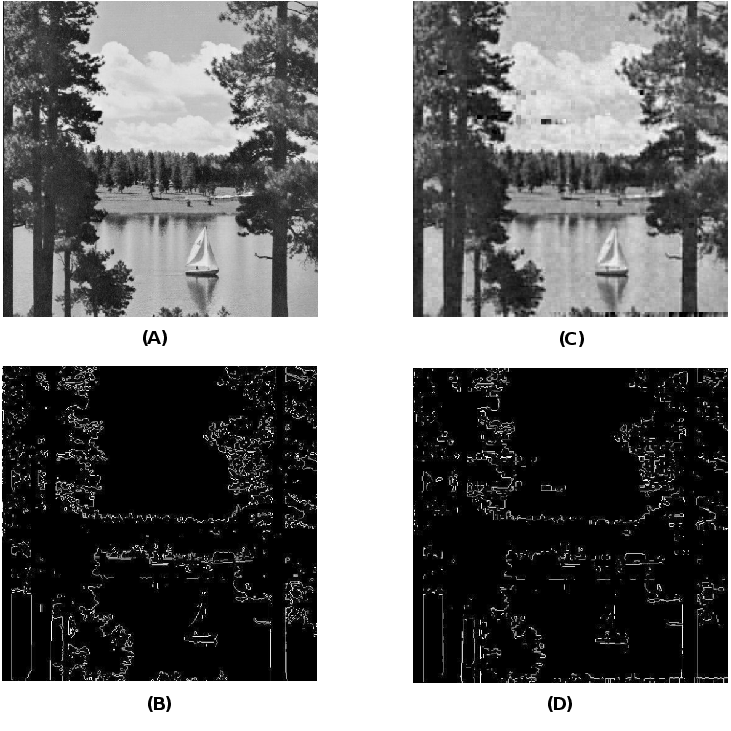}
		\caption{Visual quality analysis between the `Lake' original secret image and the `Lake' extracted secret image (from the `Stream' stego-image). (A) ‘Lake’ Original Secret Image, (B) Original Secret Image edge map, (C) ‘Lake’ Extracted Secret Image, and (D) Extracted Secret Image edge map.}
		\label{Figure:secret recovered image visual analysis SABMIS}
\end{figure}

%\begin{figure}[!h]
%	\centering
%	\begin{multicols}{2}
%	\begin{subfigure}[b]{0.4\textwidth}
%		\centering
%		\includegraphics[width=5.5cm,height=5.5cm,keepaspectratio]{lake.eps}
%		\caption{`Lake' Original Secret Image (OSI)}
%		\label{fig:lake secret image SABMIS}     
%	\end{subfigure}
%	\hfill %
%	\begin{subfigure}[b]{0.4\textwidth}
%		\centering
%		\includegraphics[width=5.5cm,height=5.5cm,keepaspectratio]{LakeSecretEdgeMap_st2.eps}
%		\caption{OSI Edge Map}
%		\label{fig:lake secret edge map SABMIS}
%	\end{subfigure}
%	\hfill %
%	\begin{subfigure}[b]{0.40\textwidth}
%		\centering
%		\includegraphics[width=5.5cm,height=5.5cm,keepaspectratio]{LakeSecretRecover_st2.eps}
%		\caption{`Lake' Extracted Secret Image (ESI)}
%		\label{fig:lake secret recovered image SABMIS}
%	\end{subfigure}
%	\hfill %
%	\begin{subfigure}[b]{0.40\textwidth}
%		\centering
%		\includegraphics[width=5.5cm,height=5.5cm,keepaspectratio]{LakeSecretRecoverEdgeMap_st2.eps}
%		\caption{ESI Edge Map}
%		\label{fig:lake secret recovered edge map SABMIS}
%	\end{subfigure}
%	\end{multicols}
%	\caption{Visual quality analysis between the `Lake' original secret image and the `Lake' extracted secret image (from the `Stream' stego-image).}
%		\label{Figure:secret recovered image visual analysis SABMIS}
%\end{figure}

%%%%%%%% Security Analysis %%%%%%%%%%%%%
%\subsection{Security Analysis}\label{Security Analysis}
\subsection{Security Analysis}\label{Security Analysis st2}
The SABMIS scheme is a transform domain based technique which employs an indirect embedding strategy, i.e., it does not follow the Least Significant Bits (LSB) flipping method, and hence, it is immune to statistical attacks \citep{Westfeld, PM1_steganography}.
Moreover, in the SABMIS scheme, the measurement matrix $\Phi$, and the embedding/ extraction algorithmic settings are considered as secret-keys, which are shared between the sender and the legitimate receiver. Even if the eavesdropper intercepting the stego-data becomes aware that SABMIS scheme has been used to embed a secret image, he would not know these secret keys. Hence, we achieve increased security in our proposed system.

To justify this, we extract the secret image in two ways, i.e., by using correct secret-keys and by using wrong secret-keys. Here, we embed only one secret image in a cover image although these experiments can be extended to the cases of embedding two, three or four secret images. Since the measurement matrix, which we use (random matrix having numbers with mean $0$ and standard deviation $1$) is one of the most commonly used measurement matrix and the eavesdropper {\color{myyaleblue}might be able to guess it}, we use this same measurement matrix while building wrong secret-keys. Here, we use the same dimension of this matrix as well, i.e., $p3 \times p2$. In reality, the guessed matrix size would be different from the original matrix size, which would make the extraction task of the eavesdropper more difficult. 

The algorithmic settings that we use will be completely unknown to the eavesdropper as above. These involve using a set of cover image coefficient indices where secret image coefficients are embedded ($p1$ and $p4$) and few constants ($\alpha=0.01$, $\beta=0.1$, $\gamma=1$ and $c=6$). While building wrong secret-keys, without changing the indices (i.e., same $p1$ and $p4$), we take the common guess of one for all constants (i.e., $\alpha=1$, $\beta=1$, $\gamma=1$ and $c=1$). In reality, the eavesdropper would not be able to correctly guess these indices as well, resulting in further challenges during extraction.

In Figure \ref{Figure:secret recovered (using correct and wrong secret key) image visual analysis SABMIS} (A) and (B), we compare the `Lake' secret image when extracted using correct and wrong secret-keys (from the `Stream' stego-image), respectively. From this figure, we see that when using correct secret-keys, the visual distortion in the extracted secret image is negligible (as evident by comparing with Figure 6(E)), and when using the wrong secret-keys, the distortion in the extracted secret image is very high (it is almost black).

\begin{figure}[!h]
	\centering
		\includegraphics[width=\textwidth]{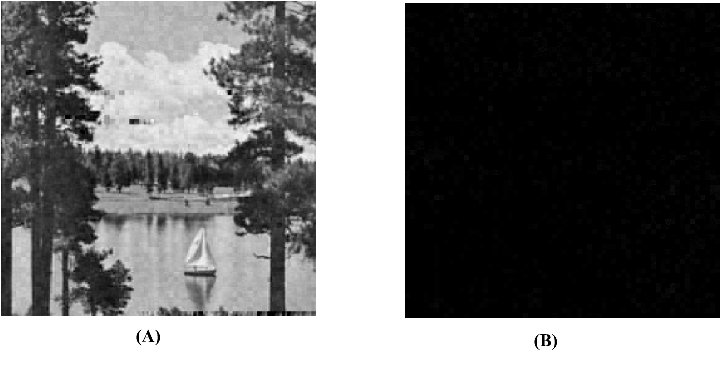}
		\caption{Visual quality analysis between the `Lake' extracted secret image using correct and wrong secret-keys (from the `Stream' stego-image). (A) ‘Lake’ extracted secret image
(using correct secret-keys), and (B) ‘Lake’ extracted secret image (using wrong secret-keys).}
		\label{Figure:secret recovered (using correct and wrong secret key) image visual analysis SABMIS}
\end{figure}

{\color{myviolet}Further, we numerically demonstrate that the correctly and wrongly extracted secret images are very different. We compute all the earlier discussed measures, i.e., PSNR, MSSIM, NCC, Entropy, and NAE values between the correctly and wrongly extracted secret images (when all four secret images had been separately embedded in the ten cover image). The average values of all these metrics are given in Table \ref{Table:NAE with a wrong secret-key (i.e., measurement matrix)}. In this table, we observe that PSNR values are very low (recall over 30 dB are considered good). The MSSIM and NCC values are close to $0$. The entropy values of correctly and wrongly extracted secret images are far from each other. Finally, NAE values are close to $1$. Hence, two images are substantially different from each other. Therefore, in the SABMIS scheme, a change in secret-keys will lead to a shift in the accuracy between the correctly and wrongly extracted secret images, in turn, making our scheme secure.}

\begin{table}[!h]
\centering
\color{myviolet}{
\footnotesize
\caption{Average PSNR, MSSIM, NCC, Entropy, and NAE value between the correctly and wrongly extracted secret images.}\label{Table:NAE with a wrong secret-key (i.e., measurement matrix)}
\setlength{\tabcolsep}{5pt}
\begin{tabular}{|c|c|c|c|cc|c|}
\hline
\multirow{2}{*}{\textbf{\begin{tabular}[c]{@{}c@{}}Cover \\ Image\end{tabular}}} & \multirow{2}{*}{\textbf{PSNR}} & \multirow{2}{*}{\textbf{MSSIM}} & \multirow{2}{*}{\textbf{NCC}} & \multicolumn{2}{c|}{\textbf{Entropy}}                                                                                                                                                                & \multirow{2}{*}{\textbf{NAE}} \\ \cline{5-6}
                                                                                 &                                &                                 &                               & \multicolumn{1}{c|}{\textbf{\begin{tabular}[c]{@{}c@{}}Correctly \\ Extracted \\ Secret Image\end{tabular}}} & \textbf{\begin{tabular}[c]{@{}c@{}}Wrongly \\ Extracted \\ Secret Image\end{tabular}} &                               \\ \hline
Fruits                                                                             & 6.032                          & 0.0116                        & 0.0037                      & \multicolumn{1}{c|}{7.188}                                                                                   & 1.409                                                                                & 0.9952                     \\ \hline
Pepper                                                                           & 5.767                         & 0.0061                         & 0.0034                      & \multicolumn{1}{c|}{7.604}                                                                                 & 1.419                                                                              & 0.9955                      \\ \hline
Boat                                                                             & 5.760                           & 0.0070                        & 0.0030                      & \multicolumn{1}{c|}{7.546}                                                                                 & 1.324                                                                               & 0.9959                      \\ \hline
House                                                                            & 5.767                         & 0.0036                        & 0.0015                        & \multicolumn{1}{c|}{7.533}                                                                                  & 0.897                                                                                & 0.9979                        \\ \hline
Lake                                                                             & 5.767                         & 0.0083                        & 0.0044                      & \multicolumn{1}{c|}{7.534}                                                                                   & 1.587                                                                                & 0.9942                       \\ \hline
Stream                                                                           & 5.835                          & 0.0113                         & 0.0071                      & \multicolumn{1}{c|}{7.542}                                                                                 & 1.974                                                                               & 0.9910                         \\ \hline
Livingroom                                                                       & 5.775                          & 0.0078                         & 0.0039                      & \multicolumn{1}{c|}{7.544}                                                                                 & 1.521                                                                               & 0.9948                        \\ \hline
Tulips                                                                             & 5.655                         & 0.0162                        & 0.0038                      & \multicolumn{1}{c|}{7.253}                                                                                  & 1.527                                                                                & 0.9948                      \\ \hline
Airplane                                                                         & 5.762                         & 0.0074                        & 0.00322                      & \multicolumn{1}{c|}{7.533}                                                                                 & 1.385                                                                                 & 0.9956                        \\ \hline
Cameraman                                                                        & 5.780                           & 0.0054                        & 0.0025                        & \multicolumn{1}{c|}{7.531}                                                                                 & 1.151                                                                               & 0.9966                      \\ \hline
\textbf{Average}                                                                          & \textbf{5.790}                        & \textbf{0.0085}                         & \textbf{0.0037 }                    & \multicolumn{1}{c|}{\textbf{7.481}}                                                                                 & \textbf{1.419}                                                                               & \textbf{0.9952}                     \\ \hline
\end{tabular}
}
\end{table}

%%%%%%%%%%%%%%%%%Timing Data
\par
\subsection{Timing Data}\label{Taming Data}
{\color{myrosewood}The time taken by our SABMIS scheme is not of great importance here because all computations are done offline, whether it is hiding of secret images, stego-image construction, or the extraction of the secret images. However, for the sake of completeness, this data, while together hiding the four secret images in the ten cover images, is given in Table \ref{Table:Timing Data}. }

\begin{table}[!h]
\centering
\color{myrosewood}
{
\footnotesize
\caption{Timing data while embedding four secret images into different cover images.}\label{Table:Timing Data}
\setlength{\tabcolsep}{5pt}
\begin{tabular}{|c|cccc|}
\hline
\multirow{2}{*}{\textbf{Cover Image}} & \multicolumn{4}{c|}{\textbf{Run Time of Different Stages of our SABMIS Scheme (in Seconds)}}                                                                                                                                                                                                             \\ \cline{2-5} 
                                      & \multicolumn{1}{c|}{\textbf{\begin{tabular}[c]{@{}c@{}}Hiding of \\Secret Images\end{tabular}}} & \multicolumn{1}{c|}{\textbf{\begin{tabular}[c]{@{}c@{}}Stego-image \\ Construction\end{tabular}}} & \multicolumn{1}{c|}{\textbf{\begin{tabular}[c]{@{}c@{}}Secret Images \\ Extraction\end{tabular}}} & \textbf{Total Time} \\ \hline
Fruits                                  & \multicolumn{1}{c|}{8.92}                    & \multicolumn{1}{c|}{74.67}                                                                             & \multicolumn{1}{c|}{12.78}                                                                             & 96.37               \\ \hline
Pepper                                & \multicolumn{1}{c|}{8.21}                    & \multicolumn{1}{c|}{73.65}                                                                             & \multicolumn{1}{c|}{10.34}                                                                             & 92.20                \\ \hline
Boat                                  & \multicolumn{1}{c|}{8.01}                    & \multicolumn{1}{c|}{76.82}                                                                             & \multicolumn{1}{c|}{8.67}                                                                              & 93.50                \\ \hline
House                                 & \multicolumn{1}{c|}{7.98}                    & \multicolumn{1}{c|}{76.58}                                                                             & \multicolumn{1}{c|}{13.86}                                                                             & 98.42               \\ \hline
Lake                                  & \multicolumn{1}{c|}{7.99}                    & \multicolumn{1}{c|}{80.07}                                                                             & \multicolumn{1}{c|}{8.42}                                                                              & 96.48               \\ \hline
Stream                                & \multicolumn{1}{c|}{10.81}                   & \multicolumn{1}{c|}{69.81}                                                                             & \multicolumn{1}{c|}{10.24}                                                                             & 90.86               \\ \hline
Livingroom                            & \multicolumn{1}{c|}{8.13}                    & \multicolumn{1}{c|}{84.15}                                                                             & \multicolumn{1}{c|}{8.49}                                                                              & 100.77              \\ \hline
Tulips                                  & \multicolumn{1}{c|}{8.68}                    & \multicolumn{1}{c|}{81.34}                                                                             & \multicolumn{1}{c|}{9.13}                                                                              & 99.15               \\ \hline
Airplane                              & \multicolumn{1}{c|}{8.43}                    & \multicolumn{1}{c|}{80.16}                                                                             & \multicolumn{1}{c|}{8.82}                                                                              & 97.41               \\ \hline
Cameraman                             & \multicolumn{1}{c|}{8.16}                    & \multicolumn{1}{c|}{79.12}                                                                             & \multicolumn{1}{c|}{8.75}                                                                              & 96.03               \\ \hline
\textbf{Average}                               & \multicolumn{1}{c|}{\textbf{8.38}}                   & \multicolumn{1}{c|}{\textbf{77.34}}                                                                            & \multicolumn{1}{c|}{\textbf{9.83}}                                                                              & \textbf{95.55}              \\ \hline
\end{tabular}
}
\end{table}

{\color{myrosewood}It is evident that the scheme is completely executed in a few minutes. Further, hiding and the extraction steps take about the same time (which they should because of similar steps), which is 10\% of the total time. The most expensive step is stego-image construction, where the optimization problem is solved, which takes 80\% of the total time.}

%%%%%%%%%%%%%%%%% Application on Real-life data %%%%%%%
\par
\subsection{Application of Our Scheme on Real-life Data}\label{Application of our scheme on real-life data}

{\color{myyaleblue}In the two subsections below (\ref{Hiding Mammograms} and \ref{Hiding Brain Images}), we experiment on hiding mammograms and brain images  (in cases where some loss is acceptable) in nondescript cover images. Sending these images safely across the internet is useful in breast cancer and brain related disease diagnosis, respectively. For the first case, we do not have reference steganographic data to compare against, while for the second case, we do have such data.}

\subsubsection{Hiding Mammograms}\label{Hiding Mammograms}
{\color{myyaleblue}
Here, we hide one through four mammograms \citep{mammogram1, mammogram2} (see two in Figure 12(A) and Figure 12(C)) into all the cover images used in our experiments. These mammograms are freely available for research purposes. 
In Table \ref{Table:Application of our scheme on real-life data}, we present the embedding capacity and PSNR values from these experiments. As evident, we obtain good embedding capacity and average as well as maximum PSNR values. The other image comparison metrics turn out to be similar as well.

In figure \ref{Figure:Visual quality analysis between `Stream' cover image (CI) and its corresponding stego-image (SI) when four mammograms images are hidden}, we present the visual comparison for `Stream' as the cover image and the corresponding stego-image. We see that the cover and its corresponding stego-image are very similar. We get analogous results for the other images as well. We also check their edge maps (as discussed in Section \ref{Stego-image Quality Assessment st2}) and obtained good results.

Next, we assess the quality of the extracted secret mammograms. In Figures 12(A) and 12(C), we show two original mammograms, and in Figures 12(B) and Figure 12(D), we show the two respective extracted mammograms (from the `Stream' stego-image). From these figures, we observe that there is very little distortion in the extracted mammograms. We get similar results for the other two mammograms as well.
}

\begin{figure}[!h]
	\centering
		\includegraphics[width=\textwidth]{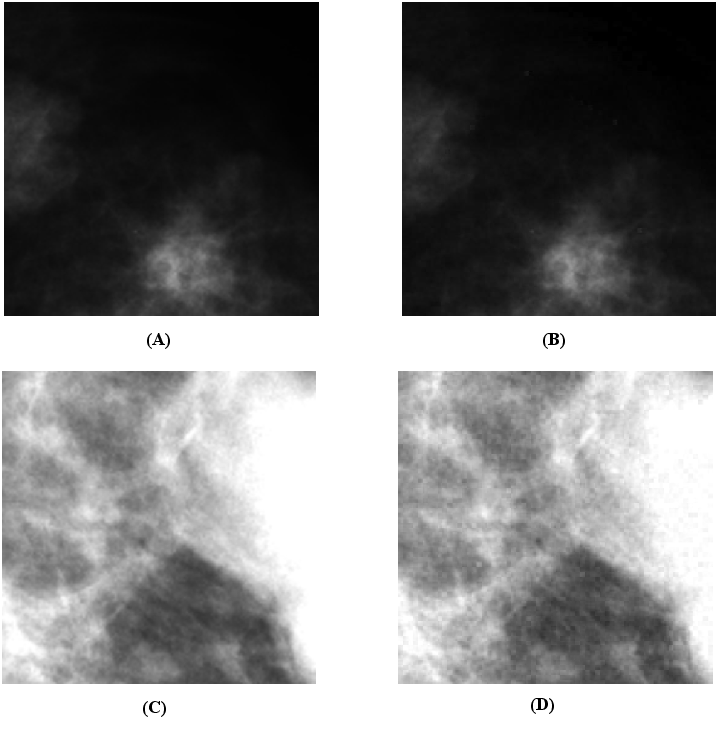}
		\caption{Visual quality analysis between the `Mammogram' original secret image and the `Mammogram' extracted secret image (from the `Stream' stego-image). (A) ‘Mammogram’ Original Secret Image \url{http://www.eng.usf.edu/cvprg/Mammography/DDSM/thumbnails/benigns/benign_14/case0355/C-0355-1.html}, (B) ‘Mammogram’ Extracted Secret Image, (C) ‘Mammogram’ Original Secret Image \url{http://www.eng.usf.edu/cvprg/Mammography/DDSM/thumbnails/cancers/cancer_15/case3517/B-3517-1.html}, and (D) ‘Mammogram’ Extracted Secret Image. Digital Database for Screening Mammography \copyright University of South Florida}
		\label{Figure:mammogram secret recovered image visual analysis SABMIS}
\end{figure}

\begin{figure}[!h]
	\centering
		\includegraphics[width=\textwidth]{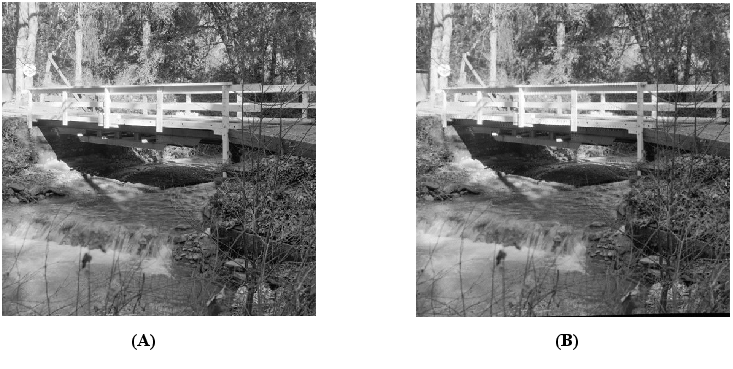}
		\caption{Visual quality analysis between `Stream' cover image (CI) and its corresponding stego-image (SI) when four mammograms are hidden. (A) Cover Image, and (B) Stego-image.}
		\label{Figure:Visual quality analysis between `Stream' cover image (CI) and its corresponding stego-image (SI) when four mammograms images are hidden}
\end{figure}

%\begin{figure}[!h]
%	\centering
%	%\begin{multicols}{1}
%	\begin{subfigure}[b]{0.4\textwidth}
%		\centering
%		\includegraphics[width=5.5cm,height=5.5cm,keepaspectratio]{MammogramHide.eps}
%		\caption{`Mammogram' Original Secret Image (OSI)}
%		\label{fig:mammogram secret image1 SABMIS}
%	\end{subfigure}
%	\begin{subfigure}[b]{0.40\textwidth}
%		\centering
%		\includegraphics[width=5.5cm,height=5.5cm,keepaspectratio]{MammogramHide_Recover.eps}
%		\caption{`Mammogram' Extracted Secret Image (ESI)}
%		\label{fig:mammogram secret recovered image1 SABMIS}
%	\end{subfigure}
%	\hfill %
%	\begin{subfigure}[b]{0.4\textwidth}
%		\centering
%		\includegraphics[width=5.5cm,height=5.5cm,keepaspectratio]{MammogramHide2.eps}
%		\caption{`Mammogram' Original Secret Image (OSI)}
%		\label{fig:mammogram secret image2 SABMIS}
%	\end{subfigure}
%	\begin{subfigure}[b]{0.40\textwidth}
%		\centering
%		\includegraphics[width=5.5cm,height=5.5cm,keepaspectratio]{MammogramHide2_Recover.eps}
%		\caption{`Mammogram' Extracted Secret Image (ESI)}
%		\label{fig:mammogram secret recovered image2 SABMIS}
%	\end{subfigure}
%	\hfill %
%%	\end{multicols}
%	\caption{Visual quality analysis between the `Mammogram' original secret image and the `Mammogram' extracted secret image (from the `Stream' stego-image).}
%		\label{Figure:mammogram secret recovered image visual analysis SABMIS}
%\end{figure}

\begin{table}[]
\centering
\footnotesize
\caption{Results of applicability of our scheme on real-life data (i.e., Mammograms).}
\label{Table:Application of our scheme on real-life data}
\setlength{\tabcolsep}{2.5pt}
\begin{tabular}{|c|c|c|c|c|c|c|}
\hline
\textbf{\begin{tabular}[c]{@{}c@{}}No. \\ of \\ secret \\ images\end{tabular}} & \textbf{\begin{tabular}[c]{@{}c@{}}Steganography \\Scheme\end{tabular}}         & \textbf{\begin{tabular}[c]{@{}c@{}}Type of \\ secret \\ image\end{tabular}} & \textbf{\begin{tabular}[c]{@{}c@{}}Type of \\ cover \\images\end{tabular}} & \textbf{\begin{tabular}[c]{@{}c@{}}EC \\ (in \si{bpp})\end{tabular}} & \textbf{\begin{tabular}[c]{@{}c@{}}(Avg. PSNR, \\ No. of \\ Cover\\ Images)\end{tabular}} & \textbf{\begin{tabular}[c]{@{}c@{}}Max. PSNR\end{tabular}}  \\ \hline
\multirow{1}{*}{1}             & {SABMIS}                       & Grayscale                                                               & Grayscale                                                                 & 2                                                  & (44.30, 10)      & 49.41                                                             \\ \hline

\multirow{1}{*}{2}             & {SABMIS}                       & Grayscale                                                               & Grayscale                                                                 & 4                                                  & (35.54, 10)      & 39.90                                                           \\ \hline

\multirow{1}{*}{3}             & {SABMIS}                       & Grayscale                                                               & Grayscale                                                                 & 6                                                  & (34.87, 10)      & 39.10                                                         \\ \hline

\multirow{1}{*}{4}             & {SABMIS}                       & Grayscale                                                               & Grayscale                                                                 & 8                                                                                  & (34.32, 10)      & 38.56                                                            \\ \hline
\end{tabular}
\end{table}

\subsubsection{Hiding Brain Images}\label{Hiding Brain Images}

{\color{myyaleblue}
The authors in \citep{S_Arunkumar_MedicalImage_DWT_2019} hide a brain image into a cover image. Since the original brain image as used in \citep{S_Arunkumar_MedicalImage_DWT_2019} is not publicly available, we work with a image that is quite similar to the image used in \citep{S_Arunkumar_MedicalImage_DWT_2019}, and is available in free public domain with no copyright (see Figure 14(A)) \citep{BrainImage1, BrainImage1Copyright}. By using SABMIS, we hide one through four copies of this image into all cover images (presented earlier), and compare with the results of \citep{S_Arunkumar_MedicalImage_DWT_2019}.

%Since the original brain image as used in \citep{S_Arunkumar_MedicalImage_DWT_2019} is not publicly available, we work with a snapshot of it from this paper (see Figure \ref{fig:brain secret image SABMIS}). By using SABMIS, we hide one through four copies of this image into all cover images (presented earlier), and compare with the results of \citep{S_Arunkumar_MedicalImage_DWT_2019}.

This comparison is given in Table \ref{Table:Application of our scheme on real-life data, and its comparison scheme with one scheme}. As evident, we are not competitive with \citep{S_Arunkumar_MedicalImage_DWT_2019} for the case of hiding one secret image (also discussed in Section \ref{Performance Comparison SABMIS}). However,  \citep{S_Arunkumar_MedicalImage_DWT_2019}\textquotesingle{s} scheme can hide only one secret image while our scheme can hide multiple secret images. We observe that using SABMIS to hide four secret images in a cover image, we obtain a good embedding capacity of \SI{8} bpp and a good average PSNR value of $33.56$. The other image comparison metrics turn out to be similar as well.

As mentioned above, \citep{S_Arunkumar_MedicalImage_DWT_2019} do not hide more than one secret image, and hence, we have no reference data to compare against in rest of our results (quality of stego-image, quality of secret image, and resistant to steganographic attacks). In Figure \ref{Figure:Visual quality analysis between `Stream' cover image (CI) and its corresponding stego-image (SI) when four medical images are hidden}, we present the visual comparison of `Stream' as the cover image and the corresponding stego-image while hiding four copies of this brain image. As evident, the cover and its corresponding stego-image are very similar. We get analogous results for the other cover images as well. We also check their edge maps (as discussed in Section \ref{Stego-image Quality Assessment st2}) and obtained good results.

In Figure \ref{Figure:medical secret recovered image visual analysis SABMIS}, we show the original brain secret image and one of the extracted brain image (from the `Stream' stego-image). From these figures, we observe that when compared with the original secret image, the quality of the extracted secret image is good. %If we closely observe, we find that the distortion in the extracted secret image when compared to the original secret image is slightly more than the distortion obtained in our main results (see Figures \ref{fig:lake secret image SABMIS} and \ref{fig:lake secret recovered image SABMIS} of Section \ref{secretImageQuality}). The reason for this is that our original secret image is not of high quality (it is just a snapshot).
Finally, like \citep{S_Arunkumar_MedicalImage_DWT_2019}, our scheme is inherently resistant to steganographic attacks. Our design makes our scheme more robust.
} 

%%%%%%%%%%%%%% Figures %%%%%%%%%%%%%
\begin{figure}[!h]
	\centering
		\includegraphics[width=\textwidth]{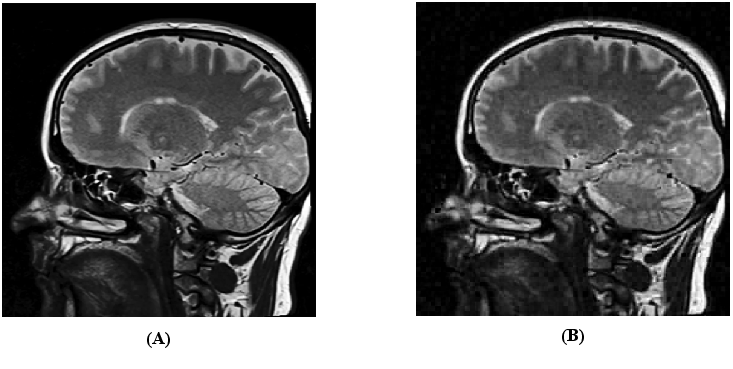}
		\caption{Visual quality analysis between the `Brain' original secret image \citep{BrainImage1, BrainImage1Copyright} and the `Brain' extracted secret image (from the `Stream' stego-image). (A) ‘Brain’ Original Secret Image \url{https://www.rawpixel.com/image/5939989/free-public-domain-cc0-photo}, and (B) ‘Brain’ Extracted Secret Image.}
		\label{Figure:medical secret recovered image visual analysis SABMIS}
\end{figure}

\begin{figure}[!h]
	\centering
		\includegraphics[width=\textwidth]{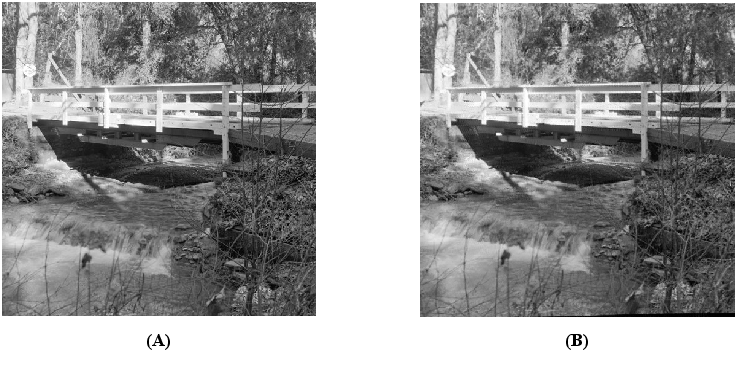}
		\caption{Visual quality analysis between `Stream' cover image (CI) and its corresponding stego-image (SI) when four copies of brain medical images are hidden. (A) Cover Image, and (B) Stego-image.}
		\label{Figure:Visual quality analysis between `Stream' cover image (CI) and its corresponding stego-image (SI) when four medical images are hidden}
\end{figure}

\begin{table}[!h]
\centering
\footnotesize
\caption{Application of our scheme on real-life data (brain image), and its comparison with one scheme.}
\label{Table:Application of our scheme on real-life data, and its comparison scheme with one scheme}
\setlength{\tabcolsep}{2.5pt}
\begin{tabular}{|c|c|c|c|c|c|c|}
\hline
\textbf{\begin{tabular}[c]{@{}c@{}}No. \\ of \\ secret \\ images\end{tabular}} & \textbf{\begin{tabular}[c]{@{}c@{}}Steganography \\Scheme\end{tabular}}         & \textbf{\begin{tabular}[c]{@{}c@{}}Type of \\ secret \\ image\end{tabular}} & \textbf{\begin{tabular}[c]{@{}c@{}}Type of \\ cover \\images\end{tabular}} & \textbf{\begin{tabular}[c]{@{}c@{}}EC \\ (in \si{bpp})\end{tabular}} & \textbf{\begin{tabular}[c]{@{}c@{}}(Avg. PSNR, \\ No. of \\ Cover\\ Images)\end{tabular}} & \textbf{\begin{tabular}[c]{@{}c@{}}Max. PSNR\end{tabular}}  \\ \hline
\multirow{1}{*}{1}                                                                 & \citep{S_Arunkumar_MedicalImage_DWT_2019} & {\color{myrosewood}Grayscale}                                                               & Grayscale                                                                    & {\color{myrosewood}2}                                                                              & (49.69, 8)            & 50.15                                                                                                                                        \\ \hline%\cline{2-8}
\multirow{1}{*}{1}                                                                      & {SABMIS}                       & Grayscale                                                               & Grayscale                                                                 & 2                                                                                  & (41.54, 10)        & 44.58                                                                                                                                 \\
        \Xhline{3\arrayrulewidth}%\hline
\multirow{1}{*}{4}                                              & {SABMIS}                       & Grayscale                                                               & Grayscale                                                                 & 8                                                                                  & (33.56, 10)         & 37.74                                                                                                                                          \\
         \Xhline{3\arrayrulewidth}%\hline
         
\end{tabular}
\end{table}

\section{Conclusions and Future Work}\label{Conclusion}
{\color{myyaleblue}In image steganography, the challenges are increasing the embedding capacity of the scheme, maintaining the quality of the stego-image as well as extracted secret image, and ensuring that the scheme is resistant to steganographic attacks.} We propose SABMIS, a blind multi-image steganography scheme for securing secret images in cover images to substantially overcome these challenges. All our images are grayscale, which is a hard problem.

{\color{myyaleblue}Our proposed SABMIS consists of many novel features to tackle the above challenges. This includes a novel embedding rule that embeds the secret image sparse coefficients into oversampled cover image sparse coefficients in a staggered manner; a transformed LASSO formulation of the underline optimization problem to construct the stego-image, which is eventually solved by ADMM; and finally, the reverse of our unique embedding rule resulting in an extraction rule. 

We perform exhaustive experiments to demonstrate that our scheme overcomes all the challenges of image steganography as discussed above. We focus on embedding multiple secret images. The embedding capacity of SABMIS for the case of embedding two and three secret images is the best in the published literature (3 times and 6 times than the existing best, respectively). While embedding four secret images, our embedding capacity is slightly lower than \citep{Ste_Ios_2006} (about ${\frac{2}{3}}^{rd}$) but we do substantially better in overcoming the other challenges.

The quality of our stego-images (when compared with the corresponding cover images) and our extracted secret images (when compared with the corresponding original secret images) are the best among the existing literature (over 30 dB of PSNR values). SABMIS is intrinsically as well as designed to be resistant to steganographic attacks (because transform based and algorithmic settings, respectively), making it one of the most secure schemes among the existing ones. 

Additionally, we show that SABMIS can be applied in very less amount of time, and also demonstrate SABMIS\textquotesingle{s} successful application on real-life problem of securely sending medical images over the internet.

Next, we discuss the future work in this context. First is further \textit{improving} our algorithm. As mentioned earlier, our SABMIS scheme has multiple novel components. Although in Appendix \ref{appendix: Sensitivity of our scheme with respect to the novel components}, we perform sensitivity analysis of SABMIS with respect to one such component (oversampling), a more detailed analysis is part of future work. In future, we plan to find improved values of parameters $\alpha, \beta, \gamma$, etc. used in the embedding and the extraction aspects of SABMIS. Further, our scheme may give poor results when embedding more than four secret images (see Appendix \ref{appendix: possible scenarios where our scheme can give less efficient remarks}). Hence, exploring this aspect is also part of our future work.

Second is \textit{extending} our scheme to embed images into videos because the amount of information that may be hidden in an image is limited. Third is \textit{adapting} our scheme for real-life applications. Although in this paper, we discuss use of SABMIS for securing mammograms and brain images while transmitting them over the internet, extensive experiments for this are part of our future work. Another related application is safely sharing biometric data of people over the internet. We plan to explore this aspect in future as well.
}
 
%\section*{Acknowledgement}
%This work has been supported by the DAAD in the project
%`A new passage to India' between IIT Indore and the Leibniz Universit\"at Hannover.

%\section*{Appendix}
\appendix
\section{Some steganography schemes for hiding binary secret data}\label{appendix:Some steganography schemes for hiding binary secret data}

{\color{myyaleblue}As discussed in the introduction, our focus is on hiding images into an image, and the images can be binary, grayscale, or color. Hiding binary data into images is a separate problem because the evaluation metrics for hiding images and binary data are completely different. However, for the sake of completeness, in Table \ref{Table:Some steganography schemes for hiding binary secret data}, we summarize some existing works that discuss hiding of binary data into images. These papers are sorted on the decreasing order of date of publishing.}

\begin{table}[!h]
\centering
\footnotesize
\caption{Some steganography schemes for hiding binary secret data into an image. All cover images are colored below.}
\label{Table:Some steganography schemes for hiding binary secret data}
\setlength{\tabcolsep}{2.5pt}
{\color{myyaleblue}
\begin{tabular}{|c|c|}
\hline
\textbf{Reference} & \textbf{Technique}                                                                                                                        \\ \hline
\citep{Gutub2021_2} & \begin{tabular}[c]{@{}c@{}}LSB (Least Significant Bits)\end{tabular}  \\ \hline 
\citep{Gutub2021} & \begin{tabular}[c]{@{}c@{}}LSB and DWT (Discrete Wavelet Transform)\end{tabular} \\ \hline 
\citep{Gutub2021_3} & \begin{tabular}[c]{@{}c@{}}LSB and DWT\end{tabular}  \\ \hline 
\citep{Gutub2020} & \begin{tabular}[c]{@{}c@{}}LSB\end{tabular}  \\ \hline 
\citep{Gutub2020_2} & \begin{tabular}[c]{@{}c@{}}LSB\end{tabular}  \\ \hline 
\citep{Gutub2019} & \begin{tabular}[c]{@{}c@{}}LSB\end{tabular}  \\ \hline 
\citep{Gutub2019_2} & \begin{tabular}[c]{@{}c@{}}A modified version of \\ LSB\end{tabular}  \\ \hline
\citep{Gutub2018} & \begin{tabular}[c]{@{}c@{}}LSB\end{tabular}  \\ \hline
\citep{Gutub2018_2} & \begin{tabular}[c]{@{}c@{}}LSB\end{tabular}  \\ \hline
\citep{Gutub2011} & \begin{tabular}[c]{@{}c@{}}A modified version of \\ LSB \end{tabular}  \\ \hline
\citep{Gutub2010} & \begin{tabular}[c]{@{}c@{}}A modified version of \\ LSB \end{tabular}  \\ \hline
\end{tabular}
}
\end{table}

\section{A small numerical example of our embedding process}\label{appendix:A small numerical example of working of our embedding process}

{\color{myyaleblue}Our embedding process for a small example (with $2\times2$ blocks for both the secret and cover images) is shown in Figure \ref{fig:A small numerical example of secret image embedding}. In the experiments, we show the results of hiding/ embedding up to four secret images in a cover image. However, for the sake of simplicity, here, we show the case of hiding one secret image into a cover image.}

\begin{figure}[!h]
	\centering
			\includegraphics[width=1\textwidth]{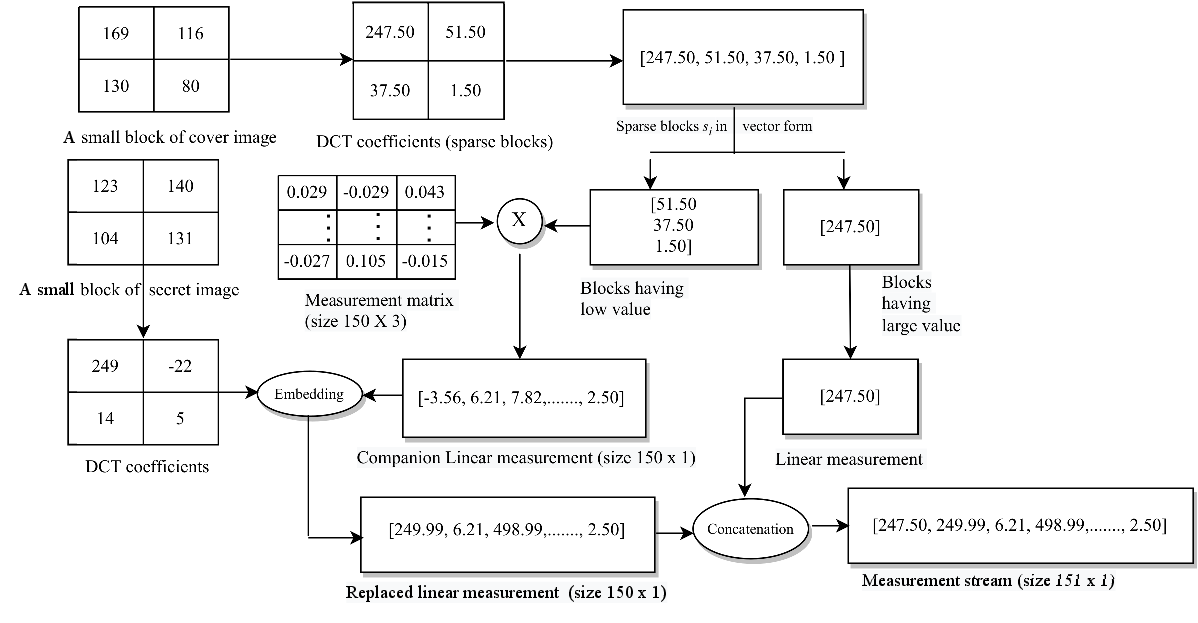}
		\caption{A small numerical example of secret image embedding.}
		\label{fig:A small numerical example of secret image embedding}
\end{figure}

\section{A small numerical example of our stego-image construction process}\label{appendix:A small numerical example of working of our stego-image construction process}
{\color{myyaleblue}Our stego-image construction process, from the stego-data obtained from Figure \ref{fig:A small numerical example of secret image embedding}, is shown in Figure \ref{fig:A small numerical example of stego-image construction}}.
%Since we consider a small example (a $2\times2$ block), there seems much difference in the block of the cover image and the stego-image. However, for the whole image, the visual difference between the original and extracted secret image is negligible, as evident from the experimental results discussed in Section \ref{Stego-image Quality Assessment st2}.

\begin{figure}[!h]
	\centering
			\includegraphics[width=1\textwidth]{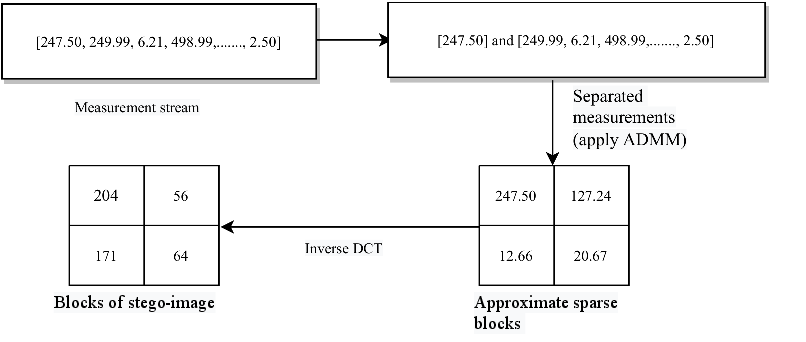}
		\caption{A small numerical example of stego-image construction.}
		\label{fig:A small numerical example of stego-image construction}
\end{figure}

\section{Sensitivity of our scheme with respect to the novel components}\label{appendix: Sensitivity of our scheme with respect to the novel components}

{\color{myviolet}
Here, we demonstrate that when we omit or restrict a particular component of our steganography scheme, then how it affects the overall performance. As discussed earlier, the novel components of SABMIS are; the oversampling of the cover image sparse coefficients and hiding secret image sparse data into them in a staggered way (our embedding rule); using ADMM to solve the LASSO formulation of the underlying minimization problem for stego-image construction; and the extraction of the secret images by the extraction rule (which is the reverse of the embedding rule). 

Without loss of generality, we restrict the oversampling component and show its effects on the performance\footnote{Since we design our embedding rule in such a way that we always need the number of linear measurements larger than the number of sparse coefficients, we could not completely omit this oversampling.}. As mentioned in the experimental result section (i.e., in Section \ref{Sec:L1SABMIS_NumericalExp}), the size of the measurement matrix $\Phi$ is $p_3\times p_2$ with $p_3 > p_2$. Earlier, we took $p_3 = 50\times p_2$. Here, we take $p_3 = 2\times p_2$, i.e., we restrict this oversampling. In Figure \ref{fig:PSNRStego1to4Imageshidden not oversampled}, we show the stego-image PSNR values for the case of hiding one, two, three, and four secret images with this restricted oversampling in SABMIS. Comparing this figure with Figure \ref{fig:PSNRStego1to4Imageshidden} (hiding one to four secret images with original oversampling in SABMIS), we observe that the PSNR values reduce substantially. Hence, the novel component of oversampling of our SABMIS scheme greatly affects the overall performance\footnote{We obtain similar results with other comparison metrics as well.}.
}

\begin{figure}[!h]
	\centering
		\includegraphics[width=1\textwidth]{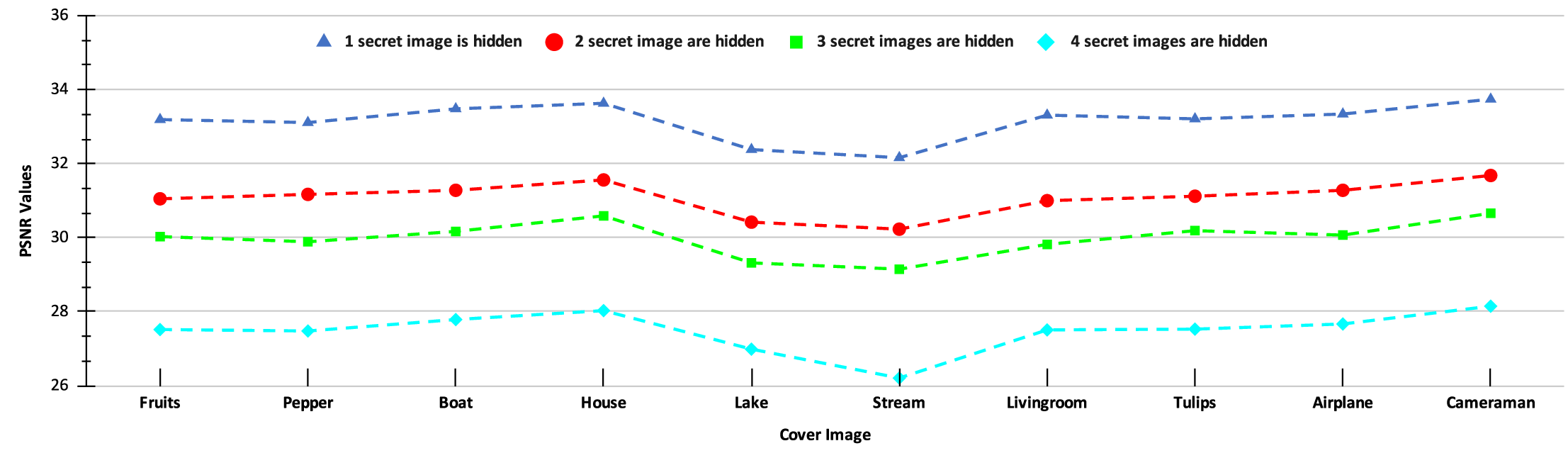}
		\caption{PSNR values of the stego-images when different numbers of images are hidden in the ten cover images (with restricted oversampling in SABMIS).}
		\label{fig:PSNRStego1to4Imageshidden not oversampled}
\end{figure}

\section{A possible scenario where our scheme is not the best}\label{appendix: possible scenarios where our scheme can give less efficient remarks}

{\color{myyaleblue}Here, we give a possible scenario where our scheme does not give the best results. We hide six (instead of four) secret images using our proposed steganography scheme and check all the evaluation metrics discussed earlier. The secret images chosen are shown in Figures 6(A), 6(B), 6(D), 6(E), 6(F), and 6(J).

We achieve up to 12 bpp embedding capacity. Visually, both the cover image and the stego-image are almost identical (see Figure \ref{Figure:visual analysis st2 hide 6}). While looking at the numerical measures, we achieve an average PSNR value of 34.39 dB, average MSSIM value close to 0.9991, average NCC value of 0.9981, nearly same entropy of the cover image and the stego-image, and average NAE value close to 0. All these values further indicate that the stego-image is very similar to its corresponding cover image. However, the original secret image and the extracted secret image are very different (see Figure \ref{Figure:visual analysis st2 secret image 6 hid}). Hence, we observe that when we try to hide more than four secret images using our scheme, the quality of extracted secret images degrades.}

\begin{figure}[!h]
	\centering
		\includegraphics[width=1\textwidth]{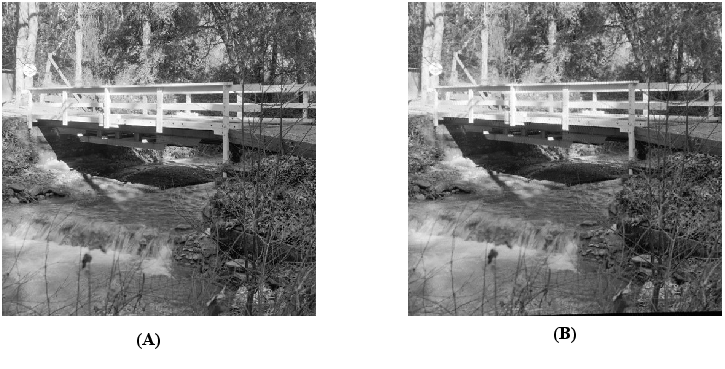}
		\caption{Visual quality analysis between `Stream' cover image (CI) and its corresponding stego-image (SI), when hiding six secret images. (A) Cover Image, and (B) Stego-image.}
	\label{Figure:visual analysis st2 hide 6}
\end{figure}

%\begin{figure}[!h]
%	\centering
%	%\begin{multicols}{1}
%	\begin{subfigure}[b]{0.40\textwidth}
%		\centering
%		\includegraphics[width=5.5cm,height=5.5cm,keepaspectratio]{Stream.eps}
%		\caption{Cover Image (CI)}
%		\label{fig:stream cover image st2 hide 6}
%	\end{subfigure}
%	\begin{subfigure}[b]{0.40\textwidth}
%		\centering
%		\includegraphics[width=5.5cm,height=5.5cm,keepaspectratio]{StegoImage_st2_stream_hid6.eps}
%		\caption{Stego-Image (SI)}
%		\label{fig:stream stego-image st2 hide 6}
%	\end{subfigure}
%	%\end{multicols}
%	\caption{Visual quality analysis between `Stream' cover image (CI) and its corresponding stego-image (SI), when hiding six secret images.}
%	\label{Figure:visual analysis st2 hide 6}
%\end{figure}

\begin{figure}[!h]
	\centering
		\includegraphics[width=1\textwidth]{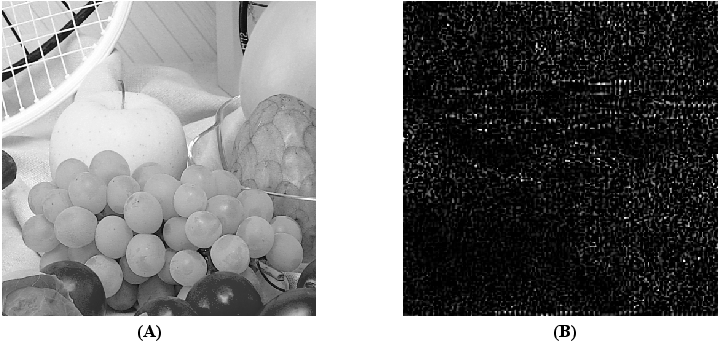}
		\caption{Visual quality analysis between the ‘Fruits’ original secret image and the ‘Fruits’ extracted secret image (from the ‘Stream’ stego-image, when hiding six secret images). (A) ‘Fruits’ Original Secret Image, and (B) `Fruits' Extracted Secret Image.}
	\label{Figure:visual analysis st2 secret image 6 hid}
\end{figure}

%\begin{figure}[!h]
%	\centering
%	%\begin{multicols}{2}
%	\begin{subfigure}[b]{0.40\textwidth}
%		\centering
%		\includegraphics[width=5.5cm,height=5.5cm,keepaspectratio]{Cars.eps}
%		\caption{‘Cars’ Original Secret Image (OSI)}
%		\label{fig:lake secret image SABMIS hide 6}
%	\end{subfigure}
%	\begin{subfigure}[b]{0.40\textwidth}
%		\centering
%		\includegraphics[width=5.5cm,height=5.5cm,keepaspectratio]{Cars-Extracted_6hid.eps}
%		\caption{‘Cars’ Extracted Secret Image (ESI)}
%		\label{fig:lake extracted secret image SABMIS hide 6}
%	\end{subfigure}
%%	\end{multicols}
%	\caption{Visual quality analysis between the ‘Cars’ original secret image and the ‘Cars’ extracted secret image (from the ‘Stream’ stego-image, when hiding six secret images).}
%	\label{Figure:visual analysis st2 secret image 6 hid}
%\end{figure}

%\bibliographystyle{ieeetr}
%\setcitestyle{square}
\bibliography{MyBibTemp}

\end{document}